\documentclass[aps,prd,10pt,twocolumn, superscriptaddress, nofootinbib]{revtex4}
\usepackage{mathrsfs}  
\usepackage{epsfig, cancel}
\usepackage{latexsym}
\usepackage{natbib, comment}
\usepackage{url}
\usepackage{dcolumn}
\usepackage{multirow}
\usepackage{color}
\usepackage{cancel}
\usepackage{soul}
\usepackage[normalem]{ulem}
\usepackage{amsfonts,amssymb,amsmath, txfonts}
\usepackage{graphicx,epsfig}
\usepackage{psfrag}
\usepackage{subfigure}
\usepackage{hyperref}
\hypersetup{colorlinks=true}
\usepackage{mathtools}
\usepackage{enumitem}
\usepackage{float}
\usepackage[dvipsnames]{xcolor}
\usepackage{xcolor}
\hypersetup{ linktoc=all,
    colorlinks, linkcolor={brightpink},
    citecolor={blue}, urlcolor={blue}
}
\definecolor{rosy}{RGB}{230,235,252}
\definecolor{myframetitle}{RGB}{90,89,170}
\definecolor{myblocktitle}{RGB}{140,185,249}
\definecolor{mytitle}{RGB}{10,80,26}

\definecolor{darkgreen}{RGB}{27,130,45}
\definecolor{darkblue}{rgb}{0,0,0.3}
\definecolor{darkred}{rgb}{0.7,0,0}

\definecolor{light gray}{RGB}{220,220,220}
\definecolor{dark purple}{RGB}{108,0,217}
\definecolor{pink}{RGB}{190,20,100}
\definecolor{orang}{RGB}{193,63,0}
\definecolor{green}{RGB}{11,98,17}
\definecolor{darkpink}{RGB}{153,0,76}
\definecolor{bluegreen}{RGB}{0,102,102}
\definecolor{greenlagan}{RGB}{0,102,0}
\definecolor{redgreen}{RGB}{102,102,0}
\definecolor{Redgreen}{RGB}{153,76,0}
\definecolor{vividviolet}{rgb}{0.62, 0.0, 1.0}
\definecolor{amaranth}{rgb}{0.9, 0.17, 0.31}
\definecolor{palatinateblue}{rgb}{0.15, 0.23, 0.89}
\definecolor{brightpink}{rgb}{1.0, 0.0, 0.5}
\definecolor{cornflowerblue}{rgb}{0.39, 0.58, 0.93}
\definecolor{deepcarminepink}{rgb}{0.94, 0.19, 0.22}
\definecolor{radicalred}{rgb}{1.0, 0.21, 0.37}

\def\H0{{\text{H}\hspace*{-2.05mm}\text{H} 0\hspace*{-1.35mm}0\ }}

\def\be{\begin{equation}}
\def\ee{\end{equation}}
\def\beq{\begin{equation}}
\def\eeq{\end{equation}}
\def\bea{\begin{eqnarray}}
\def\eea{\end{eqnarray}}
\newcommand{\dd}{\textrm{d}}

\begin{document}

\title{Can dark energy be dynamical?}

 \author{Eoin \'O Colg\'ain}\email{ocolgain@gmail.com}
 \affiliation{Center for Quantum Spacetime, Sogang University, Seoul 121-742, Korea}
 \affiliation{Department of Physics, Sogang University, Seoul 121-742, Korea} 
 \author{M. M. Sheikh-Jabbari}\email{shahin.s.jabbari@gmail.com}
\affiliation{School of Physics, Institute for Research in Fundamental Sciences (IPM), P.O.Box 19395-5531, Tehran, Iran}
\author{Lu Yin}\email{yinlu@sogang.ac.kr}
\affiliation{Center for Quantum Spacetime, Sogang University, Seoul 121-742, Korea}
 \affiliation{Department of Physics, Sogang University, Seoul 121-742, Korea}

\begin{abstract}
We highlight shortcomings of the dynamical dark energy (DDE) paradigm. For parametric models with equation of state (EOS), $w(z) = w_0 + w_a f(z)$ for a given function of redshift $f(z)$, we show that the errors in $w_a$ are sensitive to $f(z)$: if $f(z)$ increases quickly with redshift $z$, then errors in $w_a$ are smaller, and vice versa. As a result, parametric DDE models suffer from a degree of arbitrariness and focusing too much on one model runs the risk that DDE may be overlooked. In particular, we show the ubiquitous Chevallier-Polarski-Linder model is one of the least sensitive to DDE.  We also comment on  ``wiggles" in $w(z)$  uncovered in non-parametric reconstructions. Concretely, we isolate the most relevant Fourier modes in the wiggles, model them and fit them back to the original data to confirm the wiggles at $\lesssim2\sigma$. We delve into the assumptions going into the reconstruction and argue that the \textit{assumed}  correlations, which clearly influence the  wiggles, place strong constraints on field theory models of DDE. 
\end{abstract}

\maketitle

\section{Introduction}

While the physics of dark energy is obscure,  existence of a feature in data consistent with the cosmological constant $\Lambda$ across supernovae \cite{Riess:1998cb, Perlmutter:1998np}, cosmic microwave background (CMB) \cite{Aghanim:2018eyx} and baryon acoustic oscillations (BAO) \cite{Eisenstein:2005su} is compelling. The glaring inconsistency of $\Lambda$, corresponding to the dark energy equation of state (EOS) $(w=-1$), with quantum theory motivates alternative dark energy models. Starting with Quintessence \cite{Copeland:2006wr, Tsujikawa:2013fta},\footnote{See \cite{Vagnozzi:2018jhn, Banerjee:2020xcn} for a discussion on how Quintessence exacerbates Hubble tension \cite{Verde:2019ivm}.} there is now a zoo of alternative dark energy models (see \cite{Clifton:2011jh} for a review) within Effective Field Theory. Pertinently, these field theories allow for evolution in the dark energy EOS $w(z)$. This motivates a host of dynamical dark energy (DDE) parametrisations \cite{Cooray:1999da, Astier:2000as, Efstathiou:1999tm, Chevallier:2000qy, Linder:2002et, Jassal:2005qc,Barboza:2008rh} in a bid to diagnose deviations from $\Lambda$ in observational data. 

Confronted with our ignorance of $w(z)$,  one can Taylor expand $w(z)$ in redshift $z$ about its value today, $w(z) = w_0 + w_a z + O(z^2)$  \cite{Cooray:1999da, Astier:2000as}. This exercise is valid as the prevailing consensus is that dark energy is a late time, or low redshift phenomenon. Expansion in $z$ at low redshift $(z < 1)$ satisfies an obvious requirement that the expansion parameter is small \cite{Cattoen:2007sk, CH}, but this ``model", \footnote{{It is more accurately a diagnostic of DDE.}} like the Efstathiou model \cite{Efstathiou:1999tm}, is not valid at high redshift, since $w(z)$ is not bounded. This problem is solved through the celebrated Chevallier-Polarski-Linder (CPL) model \cite{Chevallier:2000qy, Linder:2002et}, which employs the other natural small number $(1-a)$, where $a=(1+z)^{-1}$ is the scale factor.  The CPL model and various alternatives \cite{Jassal:2005qc,Barboza:2008rh} are valid at high redshift and allow one to bring CMB into the DDE conversation.  See \cite{Yang:2021flj, Zheng:2021oeq} for recent studies of these models.  

Alternatively, at low redshift one can employ data reconstruction techniques to extract $w(z)$ \cite{Holsclaw:2010nb, Holsclaw:2010sk, Shafieloo:2012ht, Seikel:2012uu, Crittenden:2005wj, Crittenden:2011aa, Zhao:2017cud, Wang:2018fng}. These techniques make assumptions on the correlations, either between reconstructed data points \cite{Holsclaw:2010nb, Holsclaw:2010sk, Shafieloo:2012ht, Seikel:2012uu} or reconstructed functions, e.g. $w(z)$  \cite{Crittenden:2005wj, Crittenden:2011aa}. They have an advantage over traditional models, since the reconstruction is \textit{local}. This means that the reconstruction is more sensitive to nearby data and data farther away in redshift carries less weight.\footnote{To see this, one may plot the presumed correlation function \eqref{correlation} for different values of the width parameter $a_c$. 
See also \cite{Colgain:2021ngq} for a recent discussion on how assumptions on correlations can suppress errors in cosmological parameters.} Ultimately, these two complementary approaches have to converge if DDE is physical. Our work here highlights some issues with both approaches, begging the question, even if dark energy is dynamical, are we using the correct tools (diagnostics) to find it? 

We first make a simple observation for the  traditional parametric  approach \cite{Cooray:1999da, Astier:2000as, Efstathiou:1999tm, Chevallier:2000qy, Linder:2002et, Jassal:2005qc,Barboza:2008rh}. Namely, if one considers $w(z) = w_0 + w_a f(z)$ with $f(0)=0, f'(0)=1$, then data actually constrains the product $w_a f(z)$. This can be easily seen by Taylor expanding the Hubble diagram $H(z)$ order by order in $z$ about $z=0$ and noting that the combination $w_a f^{(n)}(z=0)$ always appears together: $w_a$ cannot be separated from $f(z)$ and its derivatives. This means that if {the data is of fairly consistent uniform quality}, $f(z)$ grows slowly with redshift, the errors on $w_a$ will be large. 

Now, recall that CPL \cite{Chevallier:2000qy, Linder:2002et} is an expansion in $(1-a)$, which is an undisputed small parameter and one shall see that it is less likely to diagnose DDE. Our observation here, which we quantify through mock realisations, echoes findings in cosmographic expansions \cite{Busti:2015xqa}. {Moreover, our observation is also in line with recent studies, e.g. FIG. 7 of \cite{Zheng:2021oeq}, where it is clear that the scale of the $w_a$ axis changes with the DDE model. Once seen, this trend may be difficult to unsee.} In contrast to CPL, the less well known Barboza-Alcaniz (BA) model \cite{Barboza:2008rh} is more likely to diagnose DDE at low redshift, while the Jassal-Bagla-Padmanabhan (JBP) model \cite{Jassal:2005qc} may make $\Lambda$ a safe bet. In short, the well known parametrisations are biased,  making it is imperative to employ a wide range of parametric DDE models in studies, e.g. \cite{Yang:2021flj, Zheng:2021oeq}.

Next we turn our attention to data reconstruction and in particular claims of wiggles in $w(z)$  \cite{Zhao:2017cud}, or its integrated density $\rho_{\textrm{de}} (z)$,  
\be
\label{density}
X(z):=\frac{\rho_{\textrm{de}} (z)}{\rho_{\textrm{de}, 0}}= \exp \left(  3 \int_0^{z} \frac{1 + w(z')}{1+z'} \dd z' \right).  
\ee
In \cite{Wang:2018fng}, however, it was adopted to use $X(z)$ instead of $w(z)$ as the independent variable. Starting from $w(z)$, it is clear from (\ref{density}) that $X(z)$ cannot change sign, while for some potentially relevant DE sectors, e.g. non-minimally coupled scalar field models, one may like to allow $X(z)$ to also change sign. That being said, it is clear from the results of  \cite{Wang:2018fng} (also \cite{Bonilla:2020wbn}) that data has a preference for $X \sim 1$, so this distinction is a little moot. 

From (\ref{density}), it is evident that wiggles in $w(z)$ around $w = -1$ translate into wiggles in $X(z)$ around $X=1$. One important input in the analysis of \cite{Zhao:2017cud, Wang:2018fng} is a constraint on correlations in the dark energy sector \cite{Crittenden:2011aa}. Importantly, the correlations are defined by two parameters, an overall normalisaton, and a parameter defining the scale beyond which correlations are suppressed. As is evident from \cite{Wang:2018fng} (appendix B), the existence (or not) of wiggles depends on the scale. A fair summary of the analysis of \cite{Wang:2018fng} may be that within the assumed correlations, there exists a parameter space where the reconstructed wiggles in $X(z)$ are favoured by Bayesian evidence over flat $\Lambda$CDM ($X=1$).  A pertinent question is then whether the correlations can be realised in a well-motivated theory, e.g. a field theory? 

Before touching upon that question, we analyse wiggles for the ``default" parameters \cite{Wang:2018fng} to ascertain if the data has an affinity for them. It should be noted that this is not quite the same range of parameters where the wiggles are favoured over flat $\Lambda$CDM by Bayesian evidence (see details in \cite{Wang:2018fng}), nevertheless, wiggles exist. By Fourier decomposing the wiggles in a given redshift range, we isolate the most relevant modes and fit them back to the original data. We find that any preference the data has for the wiggles is weak ($\lesssim 2 \sigma$), but appears to be robust. In other words, the data has a (slight) preference for wiggles. Next, by working within a field theory framework that is closely related to Quintessence, but allows excursions into the phantom regime, $w(z) < -1$, we spell out the implications of the assumptions made in \cite{Zhao:2017cud, Wang:2018fng} for a run-of-the mill field theory model. We find that the restrictions are strong at the level of field theory, which means that as data improves, these discrepancies should become transparent. The analysis, while far from conclusive, serves as an appetiser to the key question can wiggles in $w(z)$ have a field theory backend?

\section{Review of DDE} 
In this work we consider the traditional DDE \textit{parametrisations} from Table \ref{DDE_models} along with the \textit{reconstructed} $X(z)$ from Wang et al. \cite{Wang:2018fng}. As explained, it is easy to translate between $w(z)$ and $X(z)$ through equation (\ref{density}) provided $X(z) > 0$. Furthermore, this equation is robust within FLRW framework and can only breakdown in the asymptotic future ($z = -1$).\footnote{{One can find dark energy parametrisations that avoid divergences \cite{Akarsu:2015yea}.}}
\begin{table}[t]
\centering 
\begin{tabular}{c|c|c}
\rule{0pt}{3ex} Model  & $w(z)$ & $ X(z) $ \\
\hline 
\rule{0pt}{3ex} ``Redshift" \cite{Cooray:1999da, Astier:2000as} & $w_0 + w_a z $ & $ (1+z)^{3(1+w_0-w_a) } e^{3 w_a z}$ \\
\rule{0pt}{3ex} CPL \cite{Chevallier:2000qy, Linder:2002et} & $w_0 + w_a \frac{z}{1+z}$ & $(1+z)^{3(1+w_0+w_a) } e^{-\frac{3 w_a z}{1+z}}$  \\
\rule{0pt}{3ex} Efstathiou \cite{Efstathiou:1999tm} & $w_0 + w_a \ln (1+z)$ & $ (1+z)^{3(1+w_0) } e^{\frac{3}{2} w_a [\ln (1+z)]^2}$ \\
\rule{0pt}{3ex} JBP \cite{Jassal:2005qc} & $w_0 + w_a \frac{z}{(1+z)^2}$ & $ (1+z)^{3(1+w_0) } e^{\frac{3 w_a}{2} \frac{z^2}{(1+z)^2}}$ \\
\rule{0pt}{3ex} BA \cite{Barboza:2008rh} & $w_0 + w_a \frac{ z (1+z)}{1+z^2}$ & $ (1+z)^{3(1+w_0) } (1+z^2)^{{3 w_a}/{2}}$ 
\end{tabular}
\caption{DDE parametrisations/models}
\label{DDE_models}
\end{table}
To begin, let us note that neglecting the JBP \cite{Jassal:2005qc} and BA models \cite{Barboza:2008rh}, where $w(z)$ is effectively a constant beyond $z \sim 1$ (see FIG. \ref{wa}),  there is a tendency in parametric DDE models for $w(z)$ to either increase or decrease monotonically with $z$. This creates an apparent clash between the traditional DDE models and the findings of \cite{Wang:2018fng, Zhao:2017cud}. In short, if the oscillatory features in $w(z)$ or $X(z)$ reported in \cite{Wang:2018fng, Zhao:2017cud} are real, then it should be intuitively obvious that the traditional parametric DDE models will fail to detect the features, as explicitly stated elsewhere \cite{Sahni:2006pa}. We put this statement beyond doubt later.

The claims of \cite{Zhao:2017cud, Wang:2018fng} supporting a $\sim 3.7 \sigma$ preference for DDE over $\Lambda$ are intriguing.\footnote{Despite the lower $\chi^2$, as explained in \cite{Wang:2018fng}, the Bayesian evidence still favours flat $\Lambda$CDM for some specific values of parameters.} In contrast to traditional models, which build up sensitivity to the $w_a$ parameter with redshift, data reconstruction based on assumed correlations in $w(z)$ or $X(z)$, allows greater local sensitivity and in principle permits deviations from $w=-1$ ($X=1$) to be identified close to $z=0$. Once again, this can be seen from Taylor expansion by noting that $w_0:=w(z=0)$ and $\Omega_{m0}$ can only be distinguished at $O(z^2)$.  Note also that the low redshift regime is where dark energy is expected to dominate. {Remarkably, the reconstructed $X(z)$ from \cite{Wang:2018fng} has a number of wiggles in $X(z)$, some of which cannot be immediately correlated with data discrepant with Planck-$\Lambda$CDM. More precisely, there are data points that are widely recognised as being discrepant with Planck-$\Lambda$CDM \cite{Aghanim:2018eyx}, notably a high local $H_0$ \cite{Riess:2019cxk} or Lyman-alpha BAO \cite{duMasdesBourboux:2020pck}, which may buy one a wiggle or two, but additional wiggles may be an artifact of the assumptions.}

The key assumption in the line of research \cite{ Crittenden:2011aa, Zhao:2017cud} is that one can work with the correlations, 
\be
\label{correlation} 
\xi (\delta a ) := \langle [ w(a) - w^{\textrm{fid}}(a)] [ w(a') - w^{\textrm{fid}}(a')] \rangle = \frac{\xi_{w}(0)}{1+ \left( \frac{\delta a}{a_c} \right)^2 },    
\ee 
where $\delta a = a - a'$. The term on the LHS is the formal definition, whereas the expression on the RHS is how it is implemented in \cite{Crittenden:2011aa, Zhao:2017cud}. One may impose similar correlations in $X(a)$ (instead of $w(a)$) \cite{Wang:2018fng}. In this work,  we switch between correlations in $w(a)$ \cite{Zhao:2017cud} and correlations in $X(a)$ \cite{Wang:2018fng}. Here $\xi_w(0)$ denotes the normalisation factor, and as explained in \cite{Crittenden:2011aa}, $a_c$ represents a smoothing distance. Note that the denominator becomes large once $\delta a > a_c$, so correlations are suppressed beyond $a_c$. As further explained in \cite{Crittenden:2011aa}, the normalisation is related to the allowed variance, 
\be
\sigma_m^2 \approx \frac{\pi \xi_{w} (0) a_c}{a_{\textrm{max}}- a_{\textrm{min}}}, 
\ee 
and in practice the numbers $\sigma_m$ and $a_c$ are put in by hand, while $\xi_{w} (0)$ is inferred. 
The canonical values chosen in \cite{Zhao:2017cud, Wang:2018fng} are $\sigma_m = 0.04 $ and $a_c = 0.06$. {In addition, there is a prior on displacements of $X$ from $X=1$, $\Delta_X$, and the default value is $\Delta_{X} = 4$. This parameter is also  dialed and the most pronounced departure from $\Lambda$CDM was reported to happen at $\Delta_X=0.09$ \cite{Wang:2018fng}.} 

\begin{figure}[htb]
\centering
\includegraphics[width=80mm]{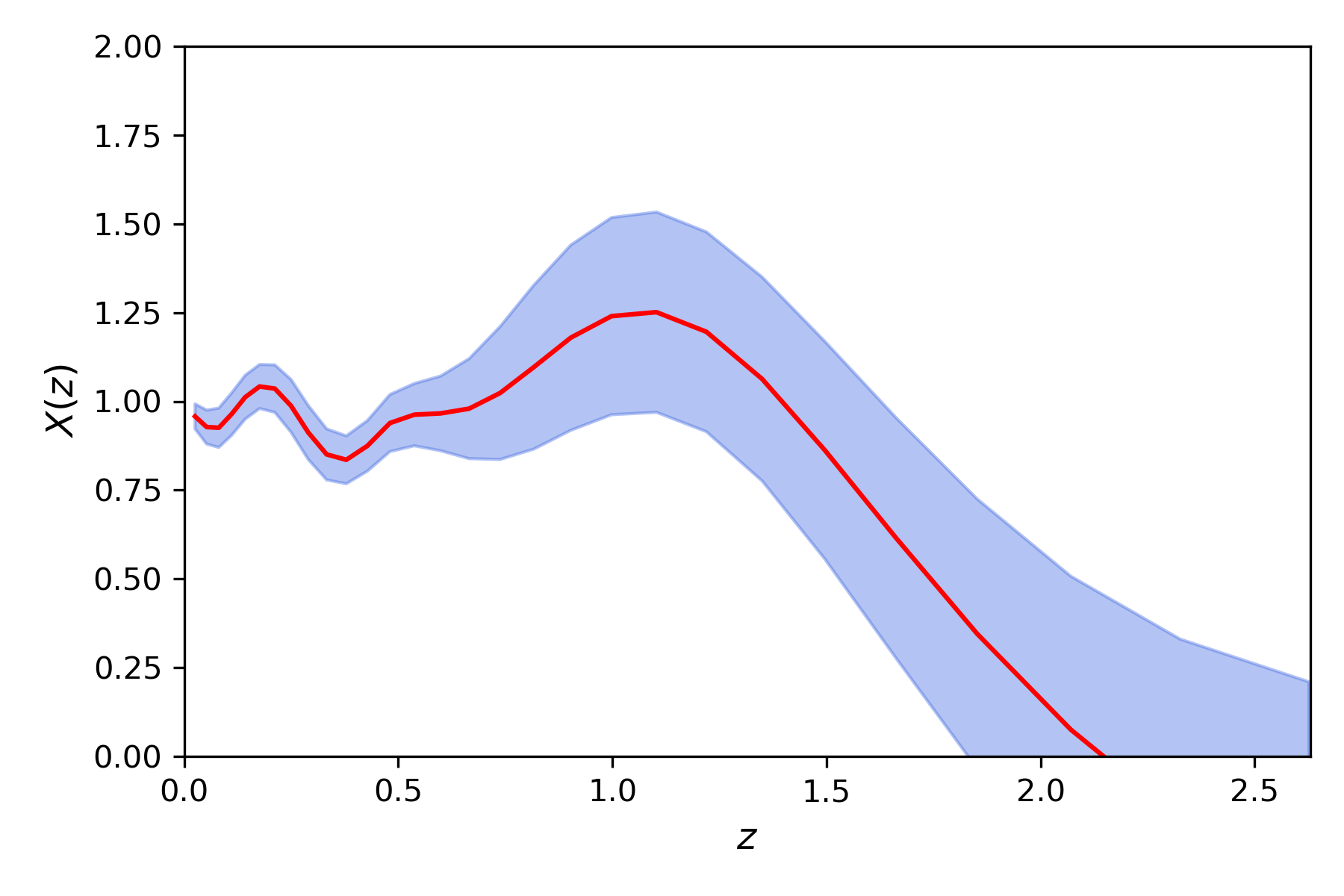} \\
\caption{Reconstructed $X(z)$ reproduced from \cite{Wang:2018fng} with parameters $\sigma_m = 0.04, a_c = 0.06, \Delta_{X} = 4$.}
\label{X_wiggle}
\end{figure}

The choice of $(\sigma_m, a_c, \Delta_{X})$ constitute transparent assumptions, and clearly, as they are dialed, one gets different results (see appendix B of \cite{Wang:2018fng}). In particular, in the limit $a_c \rightarrow 1$ or $\sigma_m \rightarrow 0$, correlations in $X(z)$ (alternatively $w(z)$) can spread further and the reconstructed function is consistent with flat $\Lambda$CDM, $X=1$ $(w=-1)$. In the later part of this work, we reanalyse the wiggles in \cite{Wang:2018fng} to ascertain if the data has a strong or weak preference for wiggles. This allows one to quantify the affinity of the data directly to wiggles without viewing them through the prism of correlations, which are objectively put in by hand.

Once the correlations (\ref{correlation}) are specified, Wang et al. \cite{Wang:2018fng} consider the Hubble parameter 
\be
\label{XCDM}
H(z) = H_0 \sqrt{ X(z) (1-\Omega_{m0} - \Omega_{r0}) + \Omega_{m0} (1+z)^3 + \Omega_{r0} (1+z)^4}, 
\ee
where $H_0$ is the Hubble constant, $\Omega_{m0}$ is the matter density and $\Omega_{r0}$ denotes the radiation density. We will largely work at low redshift where $\Omega_{r0}$ can be safely neglected. As \eqref{XCDM} shows, $X(z)$ is any contribution to the budget of the universe besides pressureless matter and radiation, which can include a DE sector plus its possible interactions with other sectors. The $X(z_i)$ parameter is reconstructed from 39 redshifts $z_i \in [ 0, 1000]$, subject to an analogous $X(a)$ correlation to (\ref{correlation}) and the further requirement that $X(a=1) = 1$. As explained in \cite{Crittenden:2011aa}, the Hubble parameter and correlation are fitted in tandem to  a combination of  data comprising CMB distance information from Planck \cite{Aghanim:2018eyx}, supernovae \cite{Betoule:2014frx}, BAO \cite{Beutler:2011hx, Ross:2014qpa, Wang:2016wjr, Font-Ribera:2013wce, Delubac:2014aqe}, cosmic chronometers \cite{Moresco:2016mzx} and a local determination of $H_0$ \cite{Riess:2016jrr}. 

We have illustrated the resulting best-fit $X(z_i)$ in FIG. \ref{X_wiggle}, while the Hubble constant $H_0$ and matter density $\Omega_{m0}$ are \cite{Wang:2018fng}, 
\be
\label{H0om}
H_0 = 70.3 \pm 0.99 \textrm{ km/s/Mpc}, \quad \Omega_{m0} = 0.288 \pm 0.008. 
\ee
It is interesting to compare the value of $\Omega_{m0} h^2$ corresponding to (\ref{H0om}), $\Omega_{m0} h^2 = 0.1423 \pm 0.008$, with the Planck value, $\Omega_{m0} h^2 = 0.1430 \pm 0.0011$ \cite{Aghanim:2018eyx}. We see that the higher value of $H_0$ and lower value of $\Omega_{m0}$, when combined, are consistent with Planck values. This may not be so surprising as while observational data is sparse in the higher redshift bins, there is some input from CMB. The high $H_0$ value has been driven by a local $H_0$ prior, but recently the rational for imposing a prior on $H_0$, versus a prior on the absolute magnitude of supernovae $M_{B}$, has been called into question \cite{Benevento:2020fev, Lemos:2018smw, Camarena:2021jlr, Efstathiou:2021ocp}. 

\section{Parametric DDE} 
Parametric models recently appeared in an assessment of DDE in light of Hubble tension by Yang et al. \cite{Yang:2021flj}. In particular, therein fits to a compilation of CMB, BAO and local $H_0$ data are performed and it is concluded that \textit{``the constraints on the cosmological parameters, both free and derived, are almost unaltered by the choice of the DE parametrization"}. This conclusion may come as no surprise. First, local determinations of $H_0$ are insensitive to the cosmological model and its dark energy sector is no exception \cite{Dhawan:2020xmp}. Secondly, CMB represents an early Universe (high redshift) observable and BAO is anchored in the early Universe. On the contrary, it is commonly believed that dark energy only becomes relevant at late times. For these reasons it may be expected that CMB constraints are largely insensitive to the details of the dark energy model. In essence, the statements in \cite{Yang:2021flj} conform to the expectations. 

That being said, when one recalls the origin of the ``redshift" \cite{Cooray:1999da, Astier:2000as}  and CPL \cite{Chevallier:2000qy, Linder:2002et} models as Taylor expansions, there is a clear distinction. It is an undeniable fact that $(1-a)=z/(1+z)$ is a smaller expansion parameter than $z$ and this has direct consequences \footnote{{In \cite{Albrecht:2006um} it has been suggested that the situation can be improved in the CPL model by replacing $w_0$ with the new parameter $w_p = w_0 + (1-a_p) w_a$, where $a_p$ is the pivot point that extremises the uncertainty in $w(a)$. This is just a redefinition of the constant component of $w(a)$ and our arguments here concern the dynamical part, i. e. the part of $w(a)$ that depends on redshift, $w_a=-dw/da$. To see this, observe that one can rewrite $w(a) = w_p + w_a (a_p -a)= w_0 + w_a (1-a)$, thus making it hopefully clear that the errors in $w_a$ are not affected by the pivot.}}. In short, in any given fit to low redshift dataset, one should expect that one has to go deeper in redshift in $(1-a)$ than $z$ in order to constrain the coefficient $w_a$. Indeed, it has already been observed by Busti et al. \cite{Busti:2015xqa} that $z$-expansions perform better than $(1-a)$-expansions at low redshift, $ z \lesssim 1.4$, i.e. within the range of supernovae, when attempting to recover the flat $\Lambda$CDM model from cosmographic expansions. In particular, it was noted that the errors in the $(1-a)$-expansion were larger. Or alternatively put, precisely because $(1-a)$ is a smaller expansion parameter, one requires a higher order Taylor expansion to approximate any model (see for example Figure 9 of \cite{Yang:2019vgk}). These statements are two faces of the same coin.  

Let us try to sum up the immediate concern. The CPL model \cite{Chevallier:2000qy, Linder:2002et} is a leading parametrisation for DDE. Objectively, current data is consistent with the cosmological constant and this means that $w_0 \approx -1$ and $w_a \approx 0$. For the DDE paradigm to be credible, one has to show that $w_a \neq 0$ outside of the confidence intervals. Now, bear in mind that evidence usually requires  a $>\!\!3\sigma$ deviation. If dark energy is largely a low redshift phenomenon, then a DDE parametrisation that is sensitive to evolution in $w(z)$ at low redshift is a prerequisite. In practice, this means that the errors on $w_a$ should be small so that deviations from $w_a = 0$ can be distinguished. Note, by DDE we are not discussing deviations from $w  = -1$, but the notion that dark energy evolves with redshift, in other words that the derivative is non-zero, $w'(z) \neq 0$, while $w_0=w(0)$ is just another parameter of the DE sector. 

\begin{figure}[htb]
\centering
\includegraphics[width=88mm]{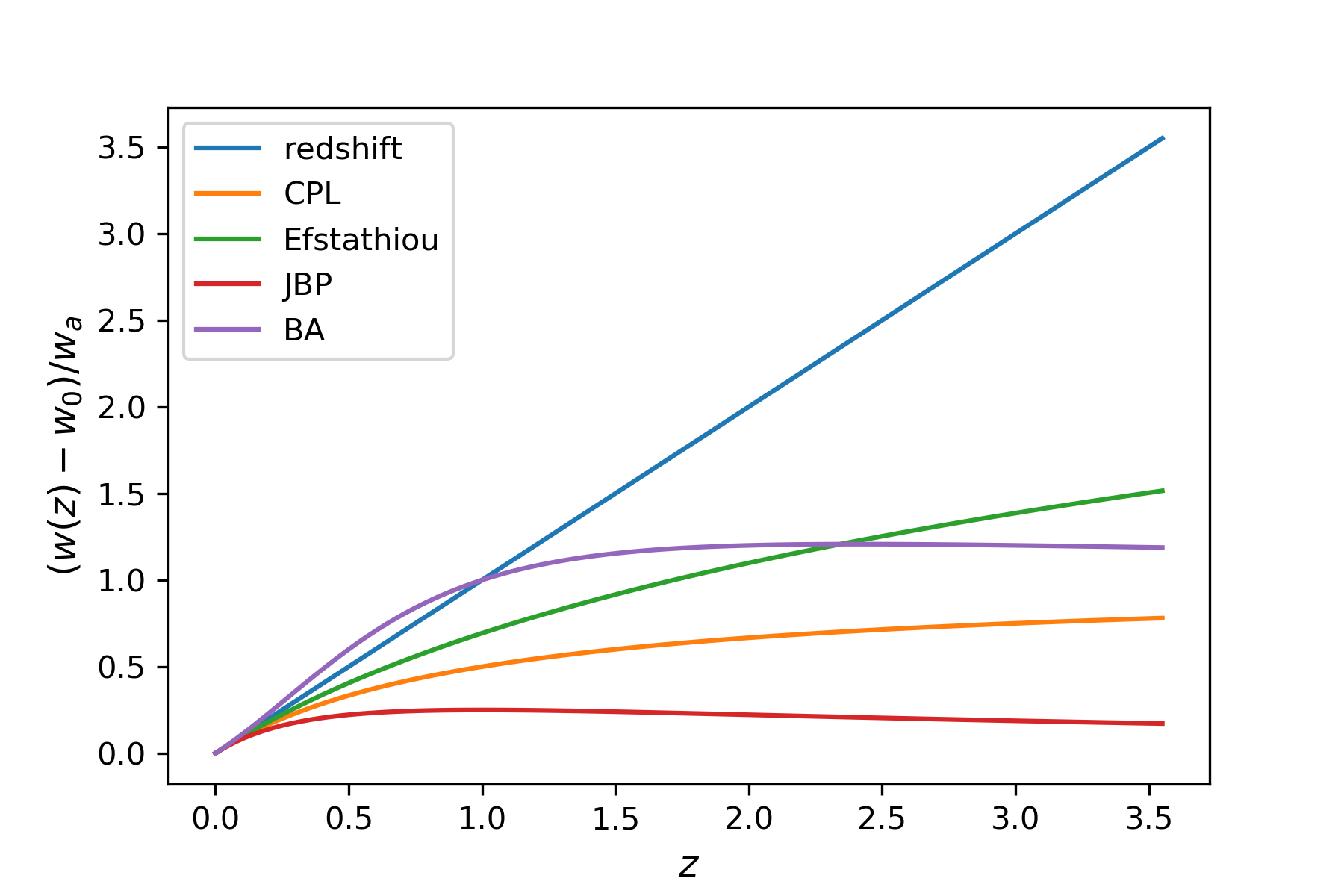} \\
\caption{The redshift dependence of various DDE models.}
\label{wa}
\end{figure}

It should be clear from the above arguments that the redshift model \cite{Cooray:1999da, Astier:2000as} will lead to smaller errors on $w_a$, thus making it a more appropriate model than the CPL model for parametrising evolution in $w(z)$ at low $z$. More generally, the errors in $w_a$ will differ across DDE models, and this leads to a degree of arbitrariness, and one of the take-home messages of this work is the necessity to analyse a number of models to reduce bias. Of course, if there is no DDE, then this arbitrariness will not be a problem. To put this comment in context, note that as we will soon show, the BA model \cite{Barboza:2008rh} leads to a detection of DDE quicker than CPL \cite{Chevallier:2000qy, Linder:2002et}, assuming DDE is real. Given how ubiquitous the CPL model has become, it is clear that some simple facts regarding these models are under-appreciated in the community. In short, parametric DDE models are biased tracers of DDE, so it is imperative to make statements across a class of models, e.g. \cite{Yang:2021flj, Zheng:2021oeq}. 

Moving along, it is easy to compare DDE models given in table \ref{DDE_models}. 
In these models 
\begin{equation}\label{wz-fz}
w(z)=w_0+ f(z)\ w_a,\qquad f(0)=0,\ \ f'(0)=1,
\end{equation}
but higher derivatives of $f(z)$ at $z=0$ differ in these models. $f(z)$ is depicted for these models in  FIG. \ref{wa}. Expanding all the $w(z)$ expressions around $z=0$, one can confirm that $w(z) \approx w_0 + w_a z $ below $z \sim 0.15$, but at higher $z$, yet still below $z \sim 1$, there are noticeable departures in behaviour. Indeed, below $z \sim 1$, provided the data is of suitably uniform quality, one can anticipate that the BA model \cite{Barboza:2008rh} should be more sensitive than the redshift model \cite{Cooray:1999da, Astier:2000as}. Furthermore, we should expect that the sensitivity to $w_a$ decreases across the  Efstathiou \cite{Efstathiou:1999tm}, CPL \cite{Chevallier:2000qy, Linder:2002et} and JBP \cite{Jassal:2005qc} models in that order.\footnote{Some of these models appeared in a recent paper \cite{Zheng:2021oeq} and as is clear from Figure 7 there, the errors in $w_a$ vary considerably. Our discussions illuminates such trends.} Below $z \sim 2$, the order in sensitivity should change so that the redshift model performs best, followed in order by BA, Efstathiou, CPL and JBP. Lastly, above $z \sim 2$ the order of sensitivity in $w_a$ changes once again and we should expect that the Efstathiou model outperforms the BA model on the size of $w_a$ errors. While this argument is analytic, and admittedly a little naive since all expressions are exact and there is no data, we will now confirm how it is realised in fits to mock data.  

\begin{figure}[htb]
\centering
\includegraphics[width=80mm]{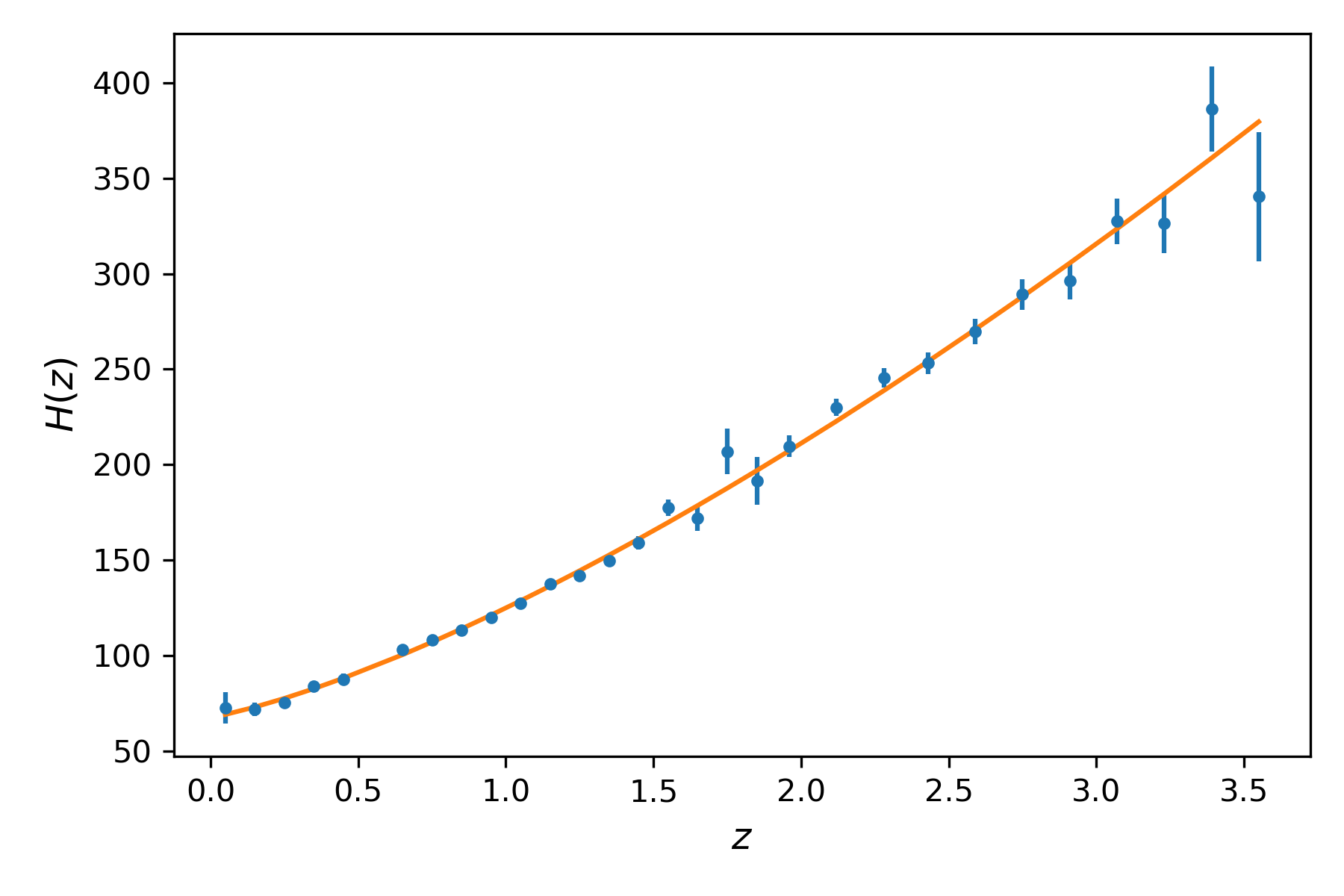}
\caption{A sample mock realisation for forecasted $H(z)$ DESI data based on the CPL model with $(H_0, \Omega_{m0}, w_0, w_a) = (67.36, 0.3153, -1, 0.5)$. }
\label{mockH}
\end{figure} 

\section{DESI Mocks} 
We begin by detailing our mocking procedure.  Since our focus is DDE, we fix the other parameters to their Planck-$\Lambda$CDM values $(H_0, \Omega_{m0}, w_0) = (67.36, 0.3153, -1)$ \cite{Aghanim:2018eyx} and choose a value of $w_a$ that is sufficiently different from $w_a = 0$. Here, we choose $w_a = 0.5$, which is clearly an exaggerated or cartoon value, but it serves to make our point. Moreover, as the focus is evolution in $w(z)$, i.e. determining $w'(z)$, it is unimportant what assumption we make on $w_0$.  The above values are nominal, but the reader is free to repeat with other values of $w_0, w_a$ and arrive at the same conclusion. Importantly, we mock data up on a particular DDE model and then fit the \textit{same} model to the mock data to recover the cosmological parameters. Note, by construction the model fits the data. We repeat this process one hundred times and average over the central values and the errors ($1 \sigma$ confidence intervals). 

\begin{figure} 
\centering
\includegraphics[width=80mm]{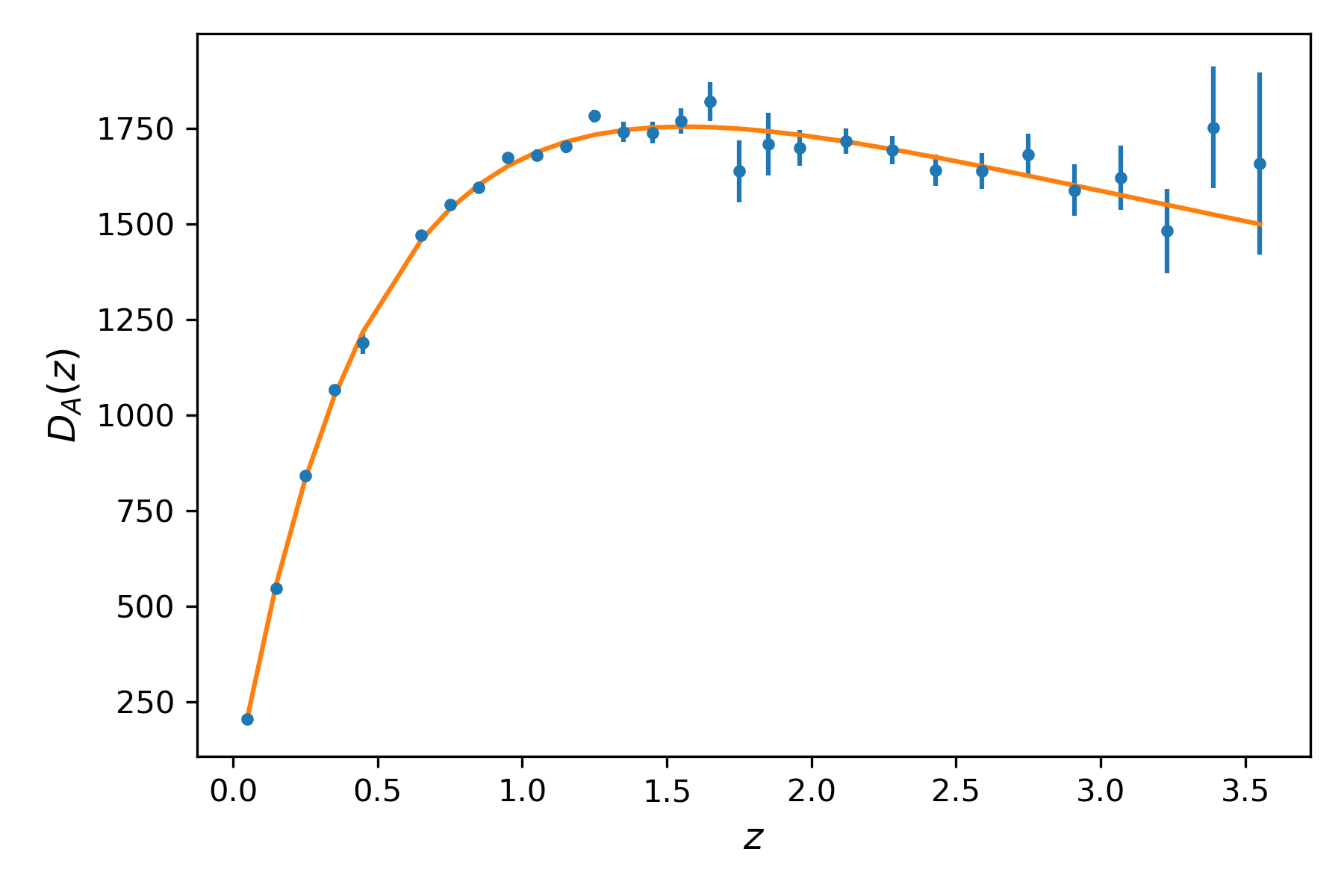}
\caption{Same as FIG. \ref{mockH} but for $D_{A}(z)$. }
\label{mockD}
\end{figure}

For the data, we use the most optimistic forecasted DESI errors on the Hubble parameter $H(z)$ and angular diameter distance $D_{A}(z)$ \cite{Aghamousa:2016zmz}  in the redshift range $0 < z \leq 3.55$, and impose a cut-off on the redshift $z_{\textrm{max}}$. We have picked this extended range so that we can flesh out as many of the features of FIG. \ref{wa} as possible. We perform Markov Chain Monte Carlo (MCMC) analysis for each realisation, and to speed up the convergence over the four parameters of interest, we impose a Planck prior $\Omega_{m0} h^2 = 0.1430 \pm 0.0011$ \cite{Aghanim:2018eyx}. We present a given mock realisation for the CPL model in FIG. \ref{mockH} and FIG. \ref{mockD}, simply to illustrate the DESI errors. We will comment on them soon.

Scanning Table \ref{table1}, one sees that with a cut-off $z_{\textrm{max}}=1$, the BA model does indeed lead to the smallest average errors on $w_a$, as anticipated from FIG. \ref{wa}. The next best performer is the redshift model, which once again confirms our expectations from FIG. \ref{wa}. Observe that, as promised, sensitivity in $w_a$ drops across the the Efstathiou, CPL and JBP models. Moreover, all of these models struggle to tell $w_a = 0.5$ apart from $w_a = 0$ with $z_{\textrm{max}}=1$. Of course, the reader can complain that $w_a = 0.5$ in the redshift model and $w_a = 0.5$ in the JBP model are different, since the combination $w_a f(z)$ is smaller in the latter, so the data will drive $w_a$ to larger values. This is true, but note that FIG. 7 of \cite{Zheng:2021oeq} uses real data and the discrepancies in the size of the $w_a$ errors are still evident (see also FIG. 2 of \cite{Barboza:2008rh}).

\begin{figure} 
\centering
\includegraphics[width=80mm]{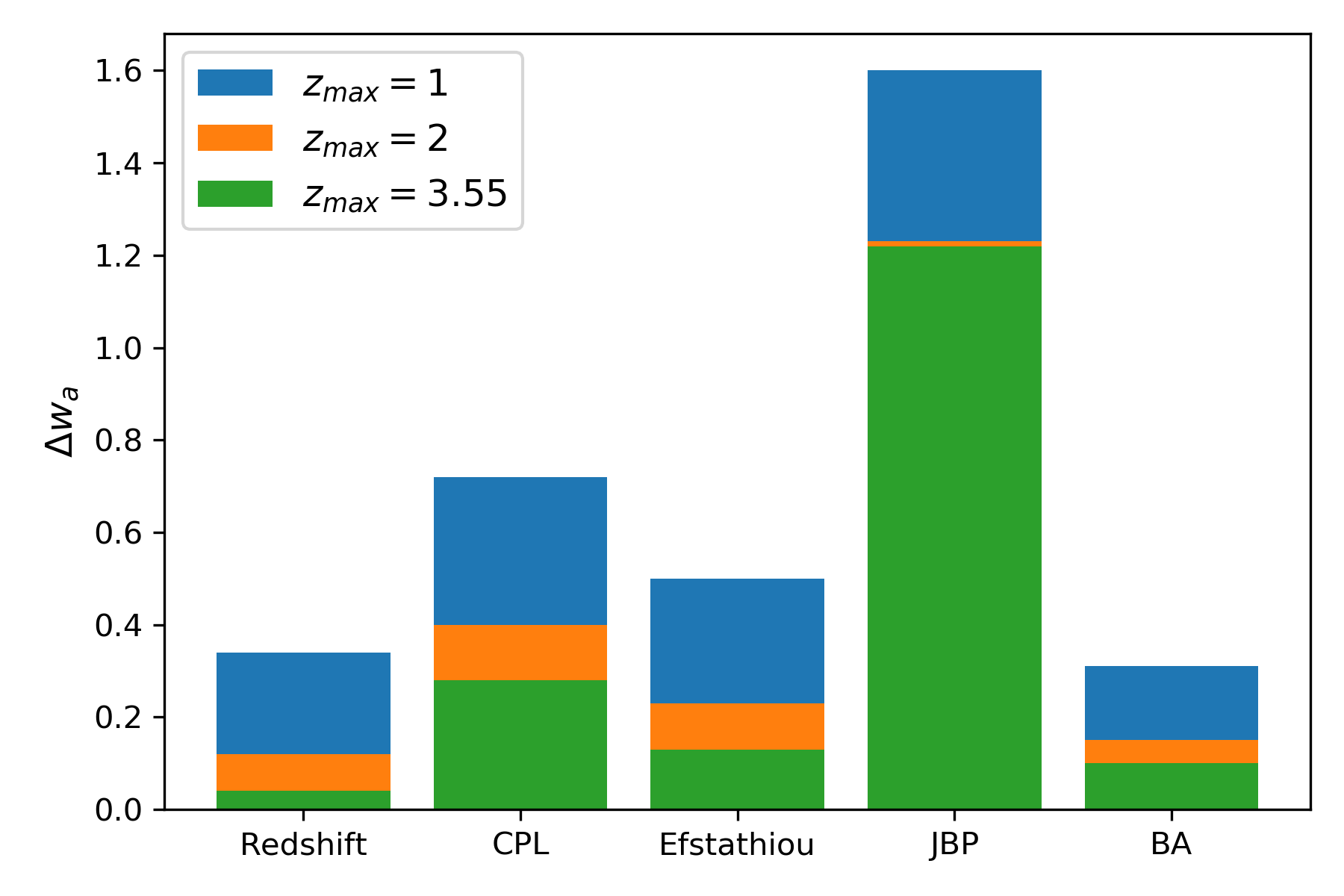}
\caption{Error in the $w_a$ parameter $\Delta w_a$ across the different models for different $z_{\textrm{max}}$ as quantified in Table \ref{table1}. }
\label{error_wa}
\end{figure}

Below $z_{\textrm{max}}=2$, the story changes, and the size of the errors in $w_a$ decrease in order across the redshift, BA, Efstathiou, CPL and JBP models. Once again this is in line with intuition gained from FIG. \ref{wa}. All the models bar JBP can now distinguish $w_a = 0.5$ from $w_a = 0$. Finally, with the highest redshift cut-off, $z_{\textrm{max}}=3.55$, the insights gleaned from FIG. \ref{wa} are largely correct, but there is a noticeable exception. From FIG. \ref{wa}, we would expect the Efstathiou model to perform better than the BA model beyond $z \sim 2.5$. However, it is clear from the numbers that this is not true. The likely explanation is that FIG. \ref{wa} is an analytic statement that does not factor in data quality. As can be seen from FIG. \ref{mockH} and FIG. \ref{mockD}, the forecasted DESI data quality is reduced at higher redshifts, so even if the Efstathiou model becomes (analytically) more sensitive to DDE than the BA model in that range, because of the decrease in data quality, this may not be evident. Note, our insights gained from analytic expressions are largely correct, but data quality plays some role. To help visualising the errors on $w_a$ with different cut-off redshifts, we plot the errors in FIG. \ref{error_wa}.  

\begin{table}[htb]
\centering 
\begin{tabular}{c|ccccc}
\rule{0pt}{3ex}  Model & $z_{\textrm{max}}$ & $H_0$ & $\Omega_{m0}$ & $w_0$ & $w_a$ \\
\hline 
\rule{0pt}{3ex} \multirow{3}{*}{Redshift} & $1$ & $67.55^{+1.53}_{-1.46}$  & $0.314^{+0.014}_{-0.014}$ & $-1.02^{+0.16}_{-0.16}$ & $0.52^{+0.34}_{-0.34}$ \\
\rule{0pt}{3ex} & $2$ & $67.20^{+0.96}_{-0.94}$  & $0.317^{+0.009}_{-0.009}$ & $-0.99^{+0.08}_{-0.07}$ & $0.48^{+0.12}_{-0.12}$ \\
\rule{0pt}{3ex} & $3.55$ & $67.38^{+0.70}_{-0.70}$  & $0.315^{+0.007}_{-0.007}$ & $-1.00^{+0.04}_{-0.04}$ & $0.50^{+0.04}_{-0.04}$  \\
\hline 
\rule{0pt}{3ex} \multirow{3}{*}{CPL} & $1$ & $67.21^{+1.74}_{-1.65}$  & $0.317^{+0.016}_{-0.016}$ & $-0.98^{+0.21}_{-0.21}$ & $0.43^{+0.72}_{-0.72}$ \\
\rule{0pt}{3ex} & $2$ & $67.39^{+1.31}_{-1.27}$  & $0.315^{+0.013}_{-0.012}$ & $-1.00^{+0.13}_{-0.13}$ & $0.49^{+0.39}_{-0.40}$ \\
\rule{0pt}{3ex} & $3.55$ & $67.18^{+1.09}_{-1.08}$ & $0.317^{+0.011}_{-0.010}$ &$-0.98^{+0.10}_{-0.10}$ & $0.46^{+0.27}_{-0.29}$ \\
\hline 
\rule{0pt}{3ex} \multirow{3}{*}{Efstathiou} & $1$ & $67.04^{+1.61}_{-1.54}$  & $0.318^{+0.015}_{-0.015}$ & $-0.96^{+0.18}_{-0.18}$ & $0.37^{+0.50}_{-0.51}$ \\
\rule{0pt}{3ex} & $2$ & $67.15^{+1.13}_{-1.10}$  & $0.317^{+0.011}_{-0.011}$ & $-0.98^{+0.10}_{-0.10}$ & $0.45^{+0.23}_{-0.23}$ \\
\rule{0pt}{3ex} & $3.55$ & $67.22^{+0.88}_{-0.86}$  & $0.317^{+0.009}_{-0.008}$ & $-0.99^{+0.07}_{-0.07}$ & $0.49^{+0.13}_{-0.13}$ \\
\hline 
\rule{0pt}{3ex} \multirow{3}{*}{JBP} &  $1$ & $67.26^{+2.06}_{-1.95}$  & $0.317^{+0.019}_{-0.019}$ & $-0.98^{+0.31}_{-0.31}$ & $0.38^{+1.59}_{-1.60}$ \\
\rule{0pt}{3ex} & $2$ & $67.27^{+1.85}_{-1.74}$  & $0.317^{+0.017}_{-0.017}$ & $-0.98^{+0.25}_{-0.25}$ & $0.40^{+1.23}_{-1.23}$ \\
\rule{0pt}{3ex} &$3.55$ & $67.40^{+1.85}_{-1.76}$  & $0.315^{+0.018}_{-0.017}$ & $-0.99^{+0.25}_{-0.25}$ & $0.43^{+1.21}_{-1.22}$ \\
\hline 
\rule{0pt}{3ex} \multirow{3}{*}{BA} & $1$ & $67.58^{+1.58}_{-1.50}$  & $0.314^{+0.015}_{-0.014}$ & $-1.01^{+0.16}_{-0.16}$ & $0.51^{+0.31}_{-0.31}$  \\
\rule{0pt}{3ex} & $2$ & $67.39^{+1.14}_{-1.11}$  & $0.315^{+0.011}_{-0.011}$ & $-1.00^{+0.10}_{-0.10}$ & $0.51^{+0.15}_{-0.15}$  \\
\rule{0pt}{3ex} & $3.55$ & $67.40^{+0.97}_{-0.95}$  & $0.315^{+0.009}_{-0.009}$ & $-1.00^{+0.075}_{-0.074}$ & $0.50^{+0.10}_{-0.10}$ 
\end{tabular}
\caption{{We show the average best-fit values of the cosmological parameters $(w_0, w_a)$ for mock DESI data with $(H_0, \Omega_{m0}, w_0, w_a) = (67.36, 0.3153, -1, 0.5)$ over 100 realisations with a redshift cut-off $z_{\textrm{max}}$.}}
\label{table1}
\end{table}

Let us summarise. As explained, parametric DDE models build up sensitivity to $w_a$ with redshift. However, the rate at which the sensitivity increases depends on the function multiplying $w_a$ in the dark energy EOS. In reality, the ubiquitous CPL model is one of the poorer performers, but not as bad as the JBP model. The CPL model \cite{Chevallier:2000qy, Linder:2002et} performs better, but is still conservative, and given how ubiquitous it has become, one may worry that not discovering DDE has become a self-fulfilling prophecy. The BA model \cite{Barboza:2008rh} performs a lot better, which should make it the parametric DDE model of choice. Of course, it still cannot recover oscillatory behaviour in $w(z)$, if it is real. 

Observe that in both the CPL and BA models, the dark energy EOS is bounded and there is no immediate obstacle to fitting CMB data. Indeed, while Barboza \& Alcaniz deserve credit for their model, and we encourage the community to use it, along with the other models to reduce bias,  the BA model may be easily tweaked it to get further improvements. To this end, note that there is a simple generalisation: 
\be
\label{genBA}
w(z) = w_0 + w_a \frac{z (1+z)^{n-1}}{1+z^n}, 
\ee
where $n \in \mathbb{N}$  and $n=1,2$ respectively correspond to the CPL and BA models. Once again, this reduces to $w(z) \approx w_0 + w_a z$ at low redshift and saturates to $w = w_0 + w_a $ at $z = \infty$, where the (generalised) BA model (\ref{genBA}) approaches the limit from above, whereas CPL approaches it from below (see FIG. \ref{wa}). In this sense, the dark energy EOS (\ref{genBA}) is on par with the CPL model, but as can be seen from the tables (for $n=2$), performs much better in constraining $w_a$. This makes it more likely that DDE, once again assuming it is physical, can be discovered. The message to the community is that one cannot rely on a single DDE parametrisation, as all of them are biased by the function $f(z)$, and it is better to study DDE over a range of models. It should be hopefully clear that if DDE is real, various models, at least in the  two-parameter $(w_0, w_a)$ family,  will not agree on the significance of any discovery. This arbitrariness will be a persistent problem, unless DDE is simply not discovered by any model!

We make one final digression to demonstrate that parametric DDE models struggle with uncovering oscillatory features in $w(z)$ or $X(z)$. Recall again that the output of the study \cite{Wang:2018fng} is the mean values of $H_0, \Omega_{m 0}, X(z_i)$ and the corresponding covariance matrix. Since the points are uniformly distributed in the scale factor $a$, but not in redshift $z$, the data points become sparse at high redshift where the only constraints come from CMB. For this reason, we restrict our attention to $z \lesssim 2.5$ (see FIG. \ref{X_wiggle}). 

Having restricted the redshift range, we crop the covariance matrix to remove the $H_0, \Omega_{m0}$ and higher redshift $X (z_i)$ entries. It is then a simple exercise to treat the remaining $X(z_i)$ as ``data" and fit the different $w(z)$ parametrisations from Table \ref{DDE_models} directly to $X (z_i)$ along with the corresponding covariance matrix. The results of the best-fit $(w_0, w_a)$ parameters are displayed in Table \ref{models}  along with their $1 \sigma$ confidence intervals. Evidently, all fits are largely consistent with the cosmological constant, i. e.  $(w_0, w_a) = (-1, 0)$ and any wiggles have been washed out. The Efstathiou model \cite{Efstathiou:1999tm} shows a small deviation from $w_0 = -1$, but this seems to be due to the fact that $w_a$ is very small. This model aside, the $w_a$ errors in Table \ref{models} are more or less in line with our expectation that the redshift and BA models are competitive, whereas the CPL and JBP models are less so. The models appear to agree on $w_a < 0$, which may be expected from the dip in $X(z)$, which is driven by Lyman-alpha BAO. So the take-home message is that if the wiggles in dark energy are real, one will not be able to probe them using traditional approaches. This appears to say that non-parametric data reconstructions have the ascendancy. 

\begin{table}[htb]
\centering 
\begin{tabular}{c|c|c}
\rule{0pt}{3ex} Model  & $w_0$ & $ w_a $ \\
\hline 
\rule{0pt}{3ex} Redshift & $-1.03^{+0.07}_{-0.07}$ & $-0.12^{+0.14}_{-0.16}$ \\
\rule{0pt}{3ex} CPL & $-1.03^{+0.09}_{-0.09}$ & $-0.19^{+0.32}_{-0.36}$ \\
\rule{0pt}{3ex} Efstathiou  & $-1.09^{+0.07}_{-0.08}$ & $-0.01^{+0.05}_{-0.05}$ \\
\rule{0pt}{3ex} JBP & $-1.05^{+0.12}_{-0.13}$ & $-0.14^{+0.77}_{-0.80}$ \\
\rule{0pt}{3ex} BA & $-1.03^{+0.08}_{-0.08}$ & $-0.10^{+0.16}_{-0.17}$ 
\end{tabular}
\caption{Direct fits of the traditional DDE models to the $X(z)$ reconstruction from \cite{Wang:2018fng}.}
\label{models}
\end{table}

\section{Discrete Fourier Transform}
\begin{figure}[htb]
\centering
  \includegraphics[width=80mm]{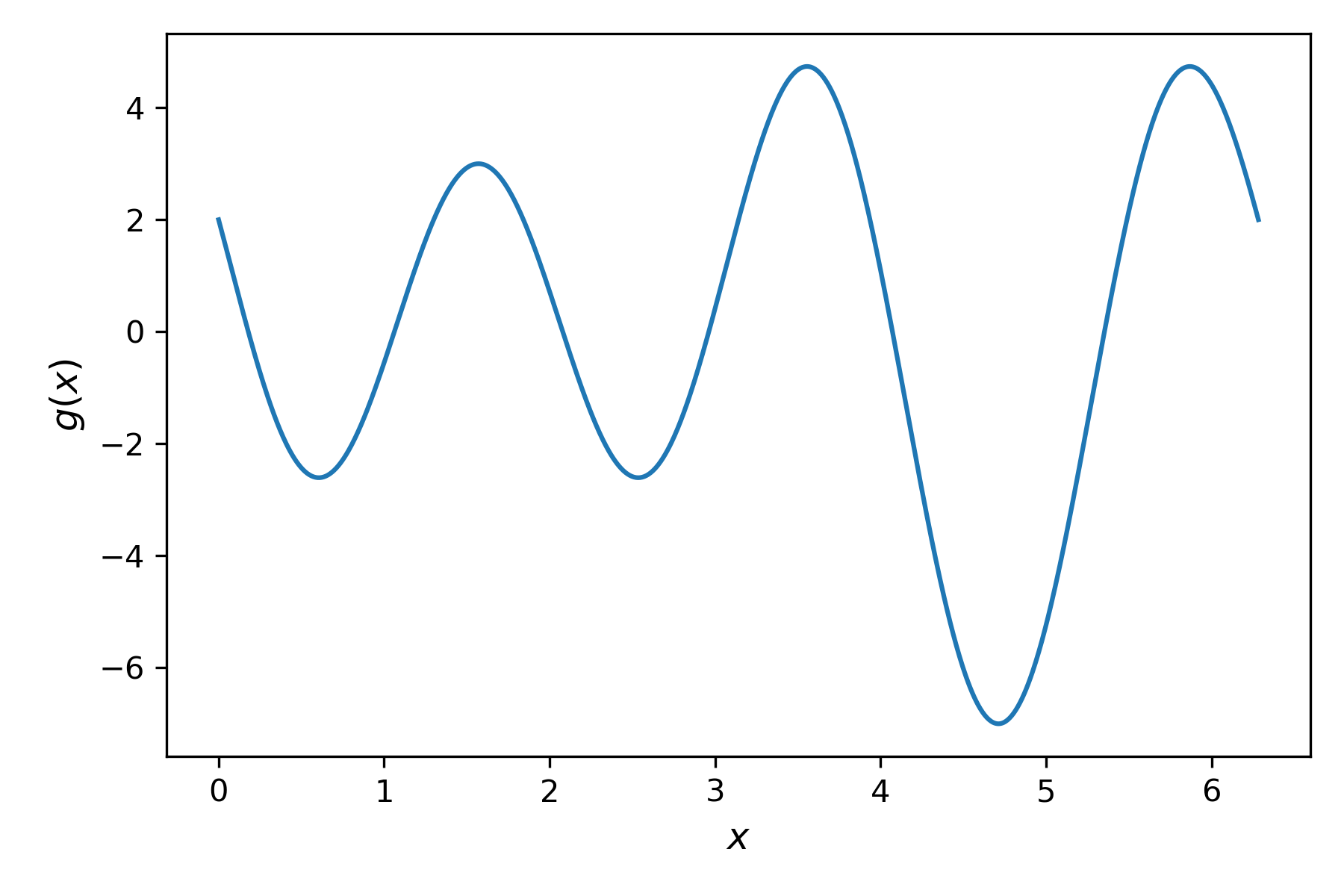}
  \caption{$g(x) =  \sin (x) + 2 \cos (2 x) - 4 \sin(3 x)$}
  \label{cos_sin}
\end{figure}

\begin{figure}
  \centering
  \includegraphics[width=80mm]{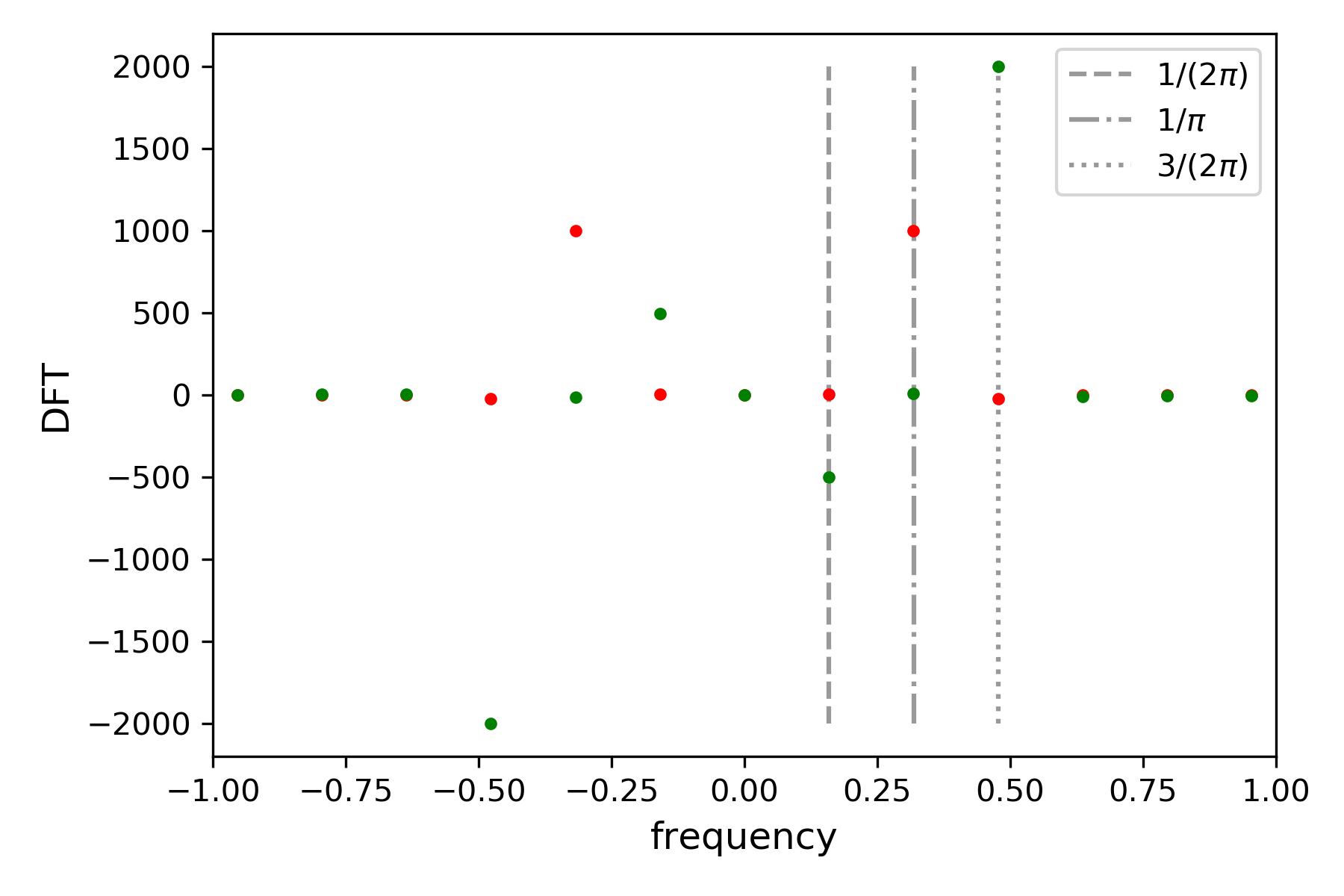}
\caption{DFT of $g(x) =  \sin (x) + 2 \cos (2 x) - 4 \sin(3 x)$ versus frequency.}
\label{cos_sin_DFT}
\end{figure}

In this section we turn our attention to the wiggles in a bid to ascertain if observational data has a preference for wiggles. This allows one to confirm or refute the output from \cite{Wang:2018fng} without resorting to the assumed correlations. In order to build a model for the wiggles in $X(z)$, we will make use of discrete Fourier transform (DFT).  To get acquainted, it is instructive to consider an example. Let  us begin with the function, 
\be
g(x) = \sin (x) + 2 \cos (2 x) - 4 \sin(3 x).  
\label{f}
\ee
We illustrate the function in FIG. \ref{cos_sin} where we have considered the period $x \in [ 0, 2 \pi]$. Noting that the curve in FIG. \ref{cos_sin} is in fact an interpolation of approximately 1000 discrete points, we have a discrete sample and one can perform a DFT analysis using the \textit{numpy.fft} package in Python.  This leads to the plot in FIG. \ref{cos_sin_DFT}. Observe that we have restricted the frequency range in the plot to the lowest frequencies of interest. In general, the DFT of a sample of discrete points is complex, so we have separated the real (red dots) and imaginary parts (green dots). 

It should be stressed that FIG. \ref{cos_sin_DFT} is specific to the function (\ref{f}) in a way that we now detail. First, most of the dots are consistent with zero, which tells us that those modes are not excited. Moreover, observe that $\sin(x), \cos(2x)$ and $\sin (3x)$ have frequencies $1/(2 \pi), 1/\pi$ and $3/(2 \pi)$, respectively, which we have highlighted using lines. In line with our expectations, dots on these lines have finite values. It should be noted that the values are also in the ratio of the coefficients, i. e. $1:2:4$, so clearly the displacement from zero encodes the amplitude of the oscillation. Finally, as is well documented, we see that the Fourier transform of a sine function is odd as the sign of the frequency is flipped, whereas the Fourier transform of cosine is even.

Having hopefully oriented ourselves, we can turn our attention to $X(z)$. The reconstructed $X(z)$ \cite{Wang:2018fng} is defined by the mean values and covariance matrix. As is clear from FIG. \ref{X_wiggle}, the dip at $z \sim 2.3$ is driven by Lyman-alpha BAO.  Neglecting the ``bump" at $z \sim 1$, which appears in a ``data desert" where there are only poor quality OHD data, the remaining wiggles are most pronounced at low redshift. In practice, we restrict our attention to $z \lesssim 0.67$ because it was easiest to approximate the wiggles using simple trigonometric functions in this redshift range. Note, there is also a little bit of trial and error here. One can construct lots of ansatze for the wiggles using the output from DFT, but one may not be able to find tangible evidence for the wiggle even in fits to $X(z)$, since the errors in $X(z)$ increase at higher redshift (see FIG. \ref{X_wiggle}). 

\begin{figure}[htb]
\centering
  \centering
  \includegraphics[width=80mm]{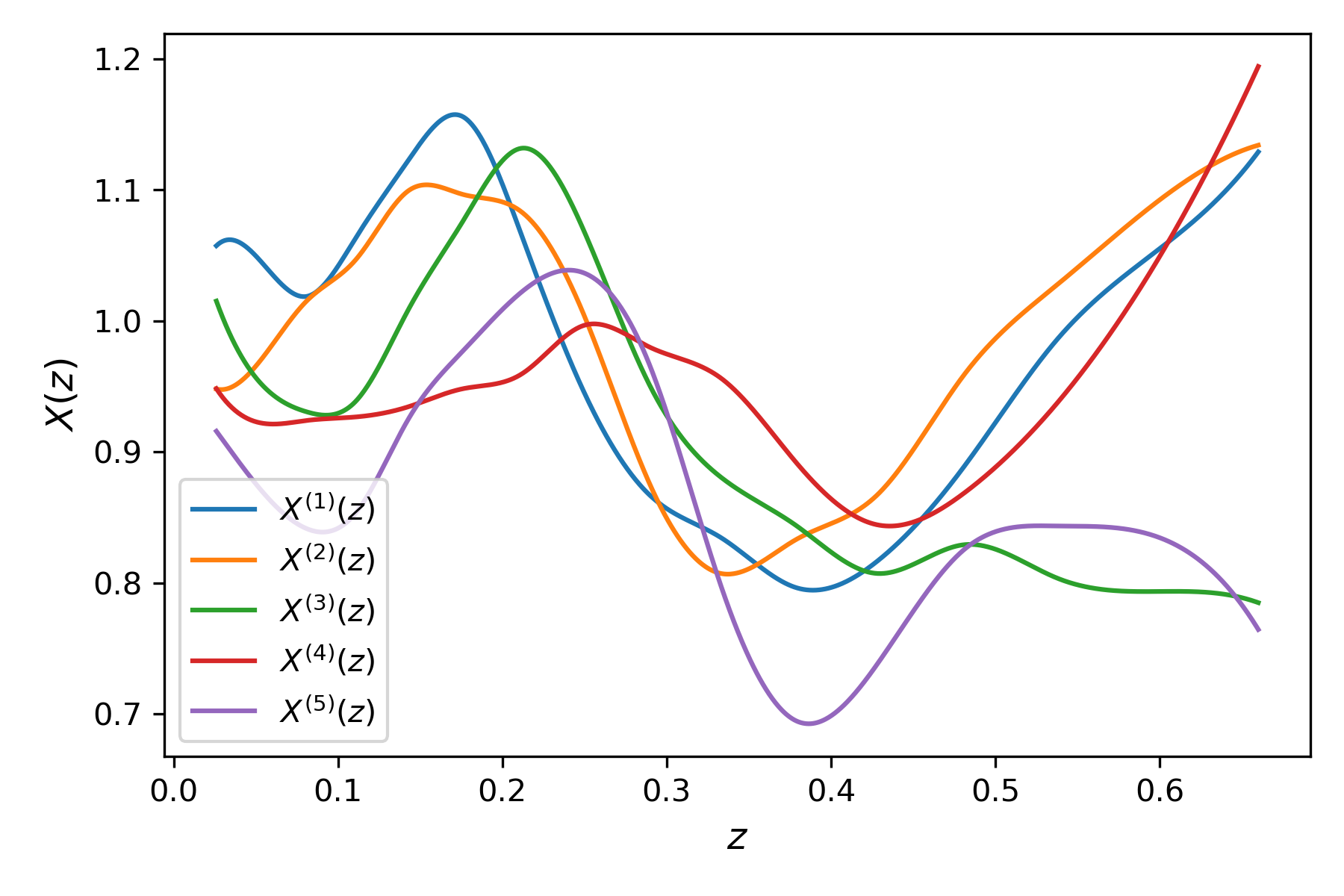}
\caption{A sample of $X(z)$ functions.}
\label{Xsample}
\end{figure}

\begin{figure} 
  \centering
  \includegraphics[width=80mm]{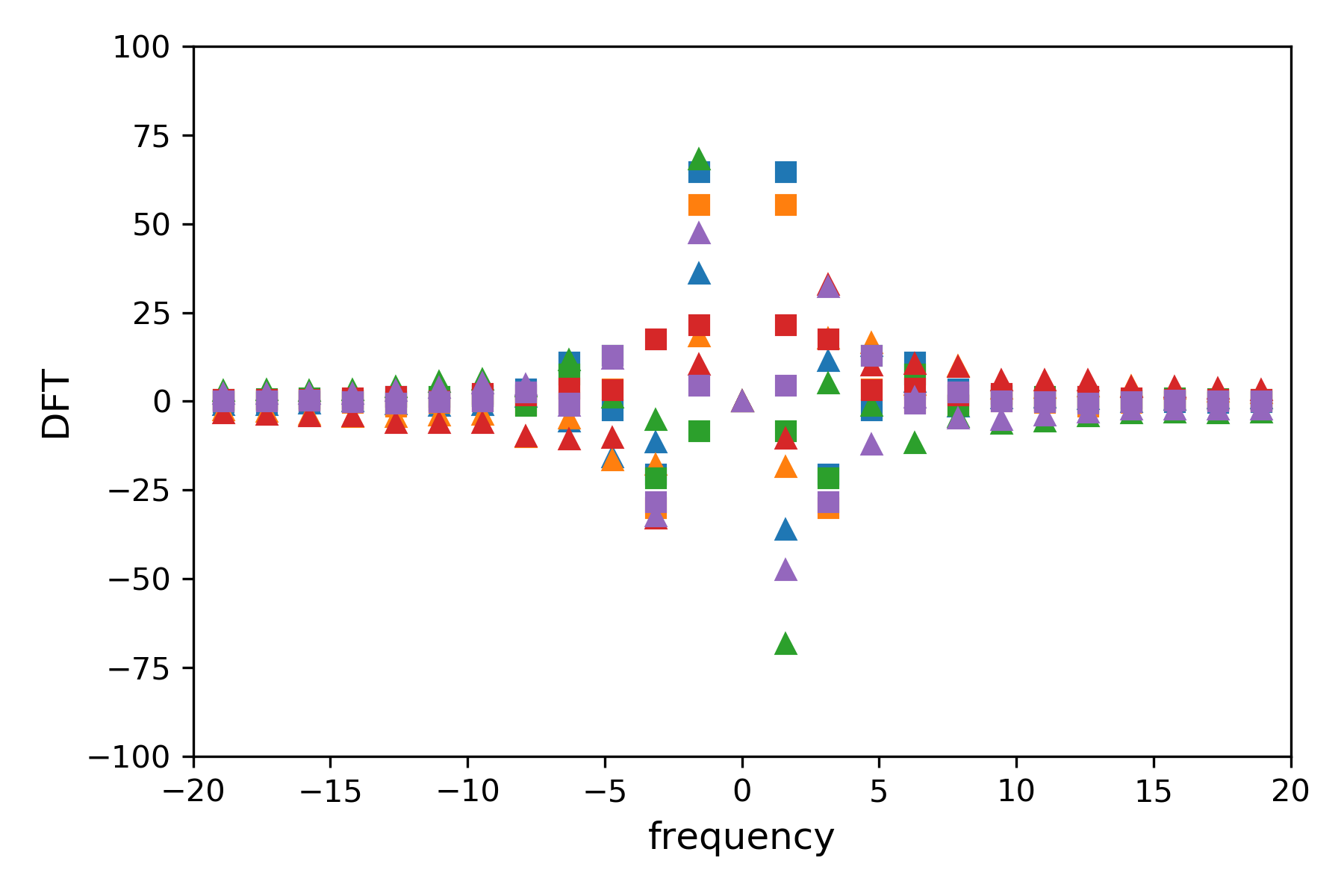}
\caption{The corresponding DFTs from the $X(z)$ sample in Fig. \ref{Xsample}.}
\label{XDFT}
\end{figure}

Now, from the mean and the covariance matrix one can generate different realisations of $X^{(n)}(z)$, a sample of which is shown in FIG. \ref{Xsample}. Performing a DFT, we arrive at the frequencies highlighted in FIG. \ref{XDFT}, where we have exploited the same colour coding as the $X^{(n)}(z)$ realisations: observe that squares denote the real components and triangles the imaginary ones. Just as in our simple warm-up example, we recognise that the higher frequency modes are largely not excited. This can be expected since the Fourier transform of the correlation (\ref{correlation}) corresponds to exponential decay. In particular, one can see that the amplitudes drop off at higher frequency and the most relevant excited frequencies $f_n$ belong to the following group, $f_n=1.58 n, n\in \mathbb{Z}$, explicitly $f_n \in \{0, \pm 1.58, \pm 3.15, \pm 4.73, \pm 6.3, \pm 7.88, \pm 9.45, \dots \}$. 

\section{Modeling the wiggles}

Having isolated the most relevant modes through DFT, let us now confirm that we have extracted the relevant modes. The first task is to check that the wiggles corresponding to the modes can be recovered from a direct MCMC exploration of the $X(z)$ reconstruction output. To do so, let us keep the lowest three modes, corresponding to the frequencies $f_1 = 1.58, f_2 = 3.15$ and $f_3 = 4.73$, which are clearly the most relevant as can be seen from FIG. \ref{XDFT}. We keep only the lowest modes, since as one increases the number of modes, the task of recovering them from an MCMC analysis becomes more daunting. Restricting the number of modes, allows us to construct the ansatz,
\be
\label{ansatz0}
X(z) = (1 - \sum_{i=1}^3 A_i) + \sum_{i=1}^3 \left[ A_i \cos ( 2 \pi f_i z) + B_i \sin (2 \pi f_i z) \right].   
\ee
Observe that although related oscillatory expressions for $w(z)$ have appeared previously in the literature \cite{Feng:2004ff, Jaime:2018ftn, Arciniega:2021ffa}, here the frequencies are fixed and the amplitudes are free parameters. Clearly, when $A_i = B_i = 0$, we recover the flat $\Lambda$CDM model where $X = 1$. This means that even in the MCMC analysis, the reconstructed $X(z)$ or original data can reject the wiggle ansatz simply by restricting $A_i$ and $B_i$ to small numbers that are consistent with zero.

\begin{figure}[htb]
\centering
  \includegraphics[width=80mm]{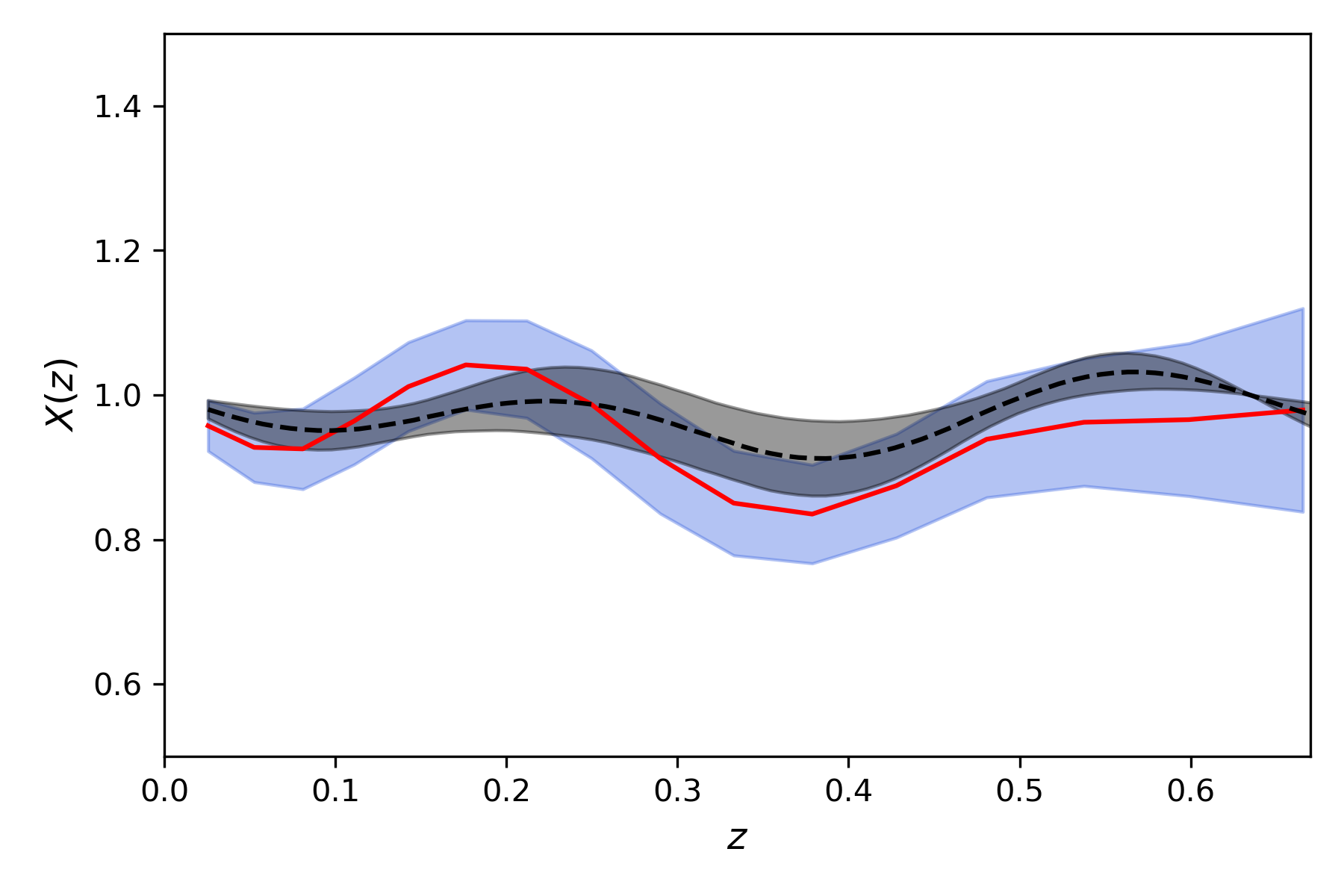}
\caption{The reconstructed $X(z)$ from \cite{Wang:2018fng} along with (\ref{ansatz2}) fitted to the same $X(z)$.}
\label{X_vs_wiggle}
\end{figure}

We first fit the general ansatz (\ref{ansatz0}) directly to the mean and covariance matrix for $X(z)$ and only retain the amplitudes that differ from zero outside of $1 \sigma$. The rational here is that those modes should stand the best chance of being recovered from the original data through further MCMC analysis. In practice, this is an iterative procedure and at each step we throw away the smallest amplitude within $1 \sigma$. This leads to the ansatz
\begin{eqnarray}
\label{ansatz1}
X(z) = (1-A_1 -A_2) &+& A_1 \cos (2 \pi f_1 z)  \\
&+& A_2 \cos (2 \pi f_2 z) + B_2 \sin (2 \pi f_2 z),  \nonumber
\end{eqnarray}
with best-fit values 
\bea
\label{fit1}
A_1 &=& 0.050^{+ 0.031}_{-0.031}, \quad A_2 = -0.030^{+ 0.026}_{-0.026}, \nonumber \\ 
B_2 &=& - 0.046^{+0.028}_{-0.027}. 
\eea
Observe that any evidence for the amplitudes is marginal. Doing the sums, one sees that $A_1, A_2$ and $B_2$ are distinct from zero at $1.6 \sigma, 1.2 \sigma$ and $1.6 \sigma$, respectively. Thus, in this context, ``evidence"  is a strong word, given that we are talking about amplitudes that differ by $\gtrsim 1 \sigma$ from zero. This may have been anticipated from FIG. \ref{X_wiggle}, as neglecting a few isolated redshift ranges, the $X=1$ line largely intersects the $1 \sigma$ confidence interval for $z \lesssim 1.6$.  Noting that evidence for $A_2$ being non-zero is more marginal than the others, one can also remove this amplitude. Doing so, we have the simplified ansatz, 
\be
\label{ansatz2}
X(z) = (1-A_1) + A_1 \cos (2 \pi f_1 z)  + B_2 \sin (2 \pi f_2 z), 
\ee
and the best-fit values become
\be
\label{fit2}
A_1 = 0.028^{+0.025}_{-0.024}, \quad B_2 = - 0.039^{+0.025}_{-0.026}, 
\ee 
which represents a deviation in $A_1$ and $B_2$ from zero of $1.2 \sigma $ and $1.6 \sigma $ statistical significance, respectively. In FIG. \ref{X_vs_wiggle} we plot the reconstructed $X(z)$ alongside the best-fit and $1 \sigma$ confidence interval inferred from the MCMC chain for our wiggle ansatz (\ref{ansatz2}). We omit a similar plot for the ansatz (\ref{ansatz1}), which since it has an extra parameter leads to a slightly broader confidence interval. It should come as no surprise to the reader that since we have truncated out the higher frequency modes, the errors have contracted noticeably. Moreover, since the ansatz (\ref{ansatz2}) is minimal, it has the smallest errors. That being said, oscillations are evident. 

We now have ansatze for a wiggle in the redshift range $z \lesssim 0.67$ and we can fit it back to a combination of low redshift data. This will establish if the wiggle is in the data or an artifact of the working assumptions employed in \cite{Wang:2018fng}. It should be noted that if the data rejects the wiggle ansatz, one should find that $A_i = B_i = 0$ within the confidence intervals. We largely make use of the same data as Wang et al. \cite{Wang:2018fng}, modulo two differences. In contrast to \cite{Wang:2018fng}, the JLA dataset \cite{Betoule:2014frx} has been replaced by Pantheon \cite{Scolnic:2017caz} and we have dropped the SH0ES prior on $H_0$ \cite{Riess:2019cxk} on the grounds that the most conservative definition of Hubble tension is that the SH0ES results require a prior on the absolute magnitude $M_B$ and not $H_0$ directly \cite{Benevento:2020fev, Lemos:2018smw, Camarena:2021jlr, Efstathiou:2021ocp}. This may look like data editing from our end, but to make the comparison fairer we will perform the analysis both with and without the priors (\ref{H0om}). By imposing the priors (\ref{H0om}), we can ensure that all constraining power in the data is being transferred to our amplitudes in the dark energy sector. This is to negate any criticism from the reader that we are fitting more parameters and it is unsurprising if we see less ``evidence" for the wiggles. 

Our best-fit values of the ansatze (\ref{ansatz1}) and (\ref{ansatz2}) to a combination of Pantheon supernovae \cite{Scolnic:2017caz}, BAO determinations from 6dF Galaxy Survey  \cite{Beutler:2011hx}, SDSS DR7 Main Galaxy Survey \cite{Ross:2014qpa}, tomographic BAO \cite{Wang:2016wjr} and cosmic chronometer data \cite{Moresco:2016mzx} can be found in Table \ref{wiggles_recovery}. Throughout we make use of the Planck prior $r_d = 147 $ Mpc and $H_0$ is in units of km/s/Mpc. Moreover, when fitting the Pantheon dataset, we made use of the following expression for the apparent magnitude $m_{B}$, $m_{B} = 5 \log_{10} (H_0 d_{L}) - 5 a_B$, where $d_{L}(z)$ is the luminosity distance and $a_B = 0.71273 \pm 0.00176$ \cite{Riess:2016jrr}. We have divided the results in Table \ref{wiggles_recovery} into the first two entries, which do not make use of the prior (\ref{H0om}), and the second two entries, which do. We have suppressed the best-fit value for $a_B$ as it is always consistent with the prior and does not add any information.  

\begin{table}[htb]
\centering 
\begin{tabular}{cccccc}
\rule{0pt}{3ex} $H_0$ & $\Omega_{m0}$ & $A_1$ & $A_2$ & $B_2$ \\
\hline 
\rule{0pt}{3ex} $68.59^{+0.56}_{-0.57}$ & $0.318^{+0.020}_{-0.020}$ & $0.012^{+0.024}_{-0.023}$ & - & $-0.026^{+0.024}_{-0.023}$  \\
\rule{0pt}{3ex} $68.66^{+0.59}_{-0.57}$ & $0.309^{+0.021}_{-0.021}$ & $0.049^{+0.031}_{-0.030}$ & $-0.050^{+0.026}_{-0.026}$ & $-0.050^{+0.028}_{-0.027}$  \\
\hline 
\rule{0pt}{3ex} $69.24^{+0.44}_{-0.42}$ & $0.291^{+0.008}_{-0.008}$ & $-0.001^{+0.017}_{-0.017}$ & - & $-0.031^{+0.023}_{-0.022}$  \\
\rule{0pt}{3ex} $69.20^{+0.45}_{-0.43}$ & $0.289^{+0.007}_{-0.008}$ & $0.045^{+0.028}_{-0.027}$ & $-0.054^{+0.026}_{-0.026}$ & $-0.055^{+0.026}_{-0.026}$  \\
\end{tabular}
\caption{Best-fit values of the models (\ref{ansatz1}) and (\ref{ansatz2}) to a compilation of supernovae, BAO and cosmic chronometer data. The upper entries do not include the prior (\ref{H0om}), whereas the lower entries do.}
\label{wiggles_recovery}
\end{table}

Now, we are in a fitting position to make a comment. The best-fit values quoted in (\ref{fit1}) and (\ref{fit2}) show fits of our wiggle ansatze directly to the reconstructed $X(z)$. That the best-fit values are non-zero outside of $1 \sigma$ confirms that the $X(z)$ ``data" recognises the wiggles. Admittedly, this recognition is marginal, nevertheless it confirms that the Fourier modes extracted from the DFT are the relevant modes. On the other hand, the results in Table \ref{wiggles_recovery} seem to confirm the presence of wiggles. There is a slight difference between the ansatze (\ref{ansatz1}) and (\ref{ansatz2}), but even without a prior (upper entries in table), the deviation from zero in the latter amplitudes are $1.6 \sigma$, $1.9 \sigma$ and $1.8 \sigma$, respectively. Once the prior is imposed this increases to $1.7 \sigma$, $2.1 \sigma$ and $2.1 \sigma$, respectively. Thus, our DFT analysis seems to confirm that the wiggles are in the data and can be accessed directly without assuming correlations. This is a consistency check and the analysis \cite{Wang:2018fng} passes convincingly.

\section{Comment on Correlations} 
While the assumptions in \cite{Wang:2018fng} are very much within the spirit of ``data-driven cosmology", if one puts correlations in by hand, and dials the parameters and finds a range where Bayesian evidence prefers the reconstructed $X(z)$ over flat $\Lambda$CDM, one may wonder what makes these values so special? Let us dwell on this further. As one sends the scale $a_c \rightarrow 1$, one recovers flat $\Lambda$CDM. There is no problem constructing large classes of field theories that allow behaviour close to flat $\Lambda$CDM. Quintessence is an example in this class. However, when one chooses $\sigma_m = 0.04, a_c = 0.06$, the presence of wiggles in $X(z)$ suggests that one may be far from the flat $\Lambda$CDM regime $(X=1)$ and it is valid to ask is there a corresponding field theory? Furthermore, if it does exist, how exotic is it? 

In this section we consider a foray into addressing this question. Let us return to (\ref{correlation}) and set $a$ or $a'$ to unity. It is then an easy task to translate from $a$ to $z$, so that the correlation may be expressed as 
\be
\xi (z) = \frac{\xi_{w}(0)}{1+ \left( \frac{z}{a_c (1+z)} \right)^2 }. 
\ee
Since we have fixed one point at $a=1$, or alternatively $z=0$, this expression now represents correlations between a given redshift $z$ and $z = 0$. Provided we operate at $z < 1$, one can Taylor expand this expression to get, 
\be
\label{xi_expand}
\xi (z) =  \xi_{w}(0) \left( 1 - \frac{z^2}{a_c^2} + \frac{2 z^3}{a_c^2} + \dots \right).  
\ee
We will now attempt to compare this to some expressions from a dynamical field theory. We make the assumption that $w^{\textrm{fid}} = -1$ and import the following expression for $w(z)$ from \cite{Banerjee:2020xcn}: 
\be 
\label{w}
w(z) = -1 + P + z Q + z^2 R + \dots 
\ee
where $\dots$ denote higher order terms and the parameters $P, Q, R$ may be expressed as,  
\begin{widetext}
\bea
\label{PQR}
P &=& \frac{\alpha^2}{3 \Omega_{\phi 0}}, \quad Q =  \frac{1}{\Omega_{\phi 0}^2} \biggl[ \frac{\alpha^4}{3} ( \Omega_{\phi 0} -1) + \frac{\alpha^2}{3} \Omega_{\phi 0} (5 - 3 \Omega_{\phi 0} ) + \frac{4}{3} \alpha \beta \Omega_{\phi 0} \biggr], \nonumber \\
R &=& \frac{1}{\Omega_{\phi 0}^3} \biggl[  \frac{\alpha^6}{6} ( \Omega_{\phi 0}^2 - 3 \Omega_{\phi 0} +2 )  + \frac{\alpha^4}{6} \Omega_{\phi 0} (17 \Omega_{\phi 0} -14 -3 \Omega_{\phi 0}^2) + 2 \alpha^3 \beta \Omega_{\phi 0} ( \Omega_{\phi 0} - 1 )   +  \frac{\alpha^2}{3} \Omega_{\phi 0}^2 (10- 9 \Omega_{\phi 0})  + \frac{4}{3} \alpha \beta \Omega_{\phi 0}^2 (5 - 3 \Omega_{\phi 0}) \nonumber \\ &+& \frac{4}{3} \beta^2 \Omega_{\phi 0}^2 + 2 \alpha \gamma \Omega_{\phi 0}^2   \biggr]. 
\eea
\end{widetext}
This expression for the EOS represents a perturbative solution to the Quintessence equations of motion about $z= 0$. Setting $\alpha = \beta = \gamma$ one recovers $w = -1$ and note that $\Omega_{\phi 0} = 1 - \Omega_{m 0}$. The parameters $\alpha, \beta, \gamma$ are related to the Quintessence potential, its first and second derivative, respectively, so even for the simplest Quintessence model one expects at least two independent parameters. It should be stressed that while Quintessence restricts $w(z) \geq -1$ by construction, this is only true for the exact solution. Since the solution outlined in \cite{Banerjee:2020xcn} is perturbative, nothing precludes values of $\alpha, \beta$ and $\gamma$ where $w < -1$. Precisely for this reason, in \cite{Banerjee:2020xcn} $w(z) \geq -1$ had to be imposed by hand.  

Now, we can make the first observation. The correlations adopted in \cite{Crittenden:2011aa, Zhao:2017cud, Wang:2018fng} tacitly assume that the dark energy sector, whether it is parametrised by $w(z)$ or $X(z)$, decouples from the other cosmological parameters $H_0$ and $\Omega_{m0}$. As can be seen from (\ref{w}), this is not necessarily true. There is a very simple reason for this. Neither $w(z)$ nor $X(z)$ are fundamental in a field theory description and both are \textit{derived} quantities. 

In addition, it is straightforward to calculate 
\bea
\xi(z)  &=&   \langle [ w(0) - w^{\textrm{fid}}(0)] [ w(z) - w^{\textrm{fid}}(z)] \rangle  \nonumber \\
&=& \langle P^2  \rangle + z \langle  P Q   \rangle  + z^2 \langle P R \rangle + \dots
\eea
where we have separated terms by $z$ dependence. Note also that, as highlighted earlier, $\Omega_{\phi 0}$ is hanging around in contradiction to assumptions made in \cite{Zhao:2017cud, Wang:2018fng}. Admittedly expressions are complicated, but this is the price one pays for working with a derived quantity. The first term is a number that can be fixed through comparison to (\ref{xi_expand}), but the term linear in $z$ must vanish. If this vanishes, this is clearly an accident and it cannot be expected to hold in general. 

Indeed, we can take this comparison a bit further and fit the original parameters $H_0, \Omega_{m0}, \alpha, \beta, \gamma$ to some representative low redshift observational data \cite{Scolnic:2017caz, Beutler:2011hx, Ross:2014qpa, Wang:2016wjr, Moresco:2016mzx}. As explained in \cite{Banerjee:2020xcn}, since the solution is perturbative, we restrict to a range of redshifts $z \lesssim 0.7$, where one can approximate the Hubble parameter to flat $\Lambda$CDM to $1 \%$ precision. This leads to an MCMC chain in the relevant parameters and using (\ref{PQR}) one can convert these chains into chains in $P, Q, R$. From there, it is easy to extract the covariance matrix and divide through by the errors to fix $\langle P^2 \rangle = \langle Q^2 \rangle = \langle R^2 \rangle  = 1$ and get an estimate for the needed correlations, e.g. $\langle  P Q   \rangle$ and $\langle P R \rangle$. This allows us to get an expression for the correlation up to an overall constant $\kappa$: 
\be
\label{xi_Quint}
\xi(z) = \kappa \left( 1 - 0.60 z + 0.21 z^2 + \dots  \right). 
\ee
Comparing this to (\ref{xi_expand}), while the overall constant is no problem, since one can tune the normalisation factor, we see the necessity of the linear term. In addition, $a_c$ can also be dialed to accommodate the quandratic term. It is worth noting that since we have switched from the scale factor $a$ to redshift $z$, $a_c = 0.06$ is no longer the relevant value and a larger value is required \cite{Raveri:2017qvt, Espejo:2018hxa}. Note also that we have adopted the prior $w^{\textrm{fid}} = -1$, but even if it is some constant value, it is difficult to see how one can remove the linear in $z$ term from (\ref{xi_Quint}) so that it agrees with (\ref{xi_expand}). In summary, the assumed correlations appear to be at odds with correlations in a representative field theory. In particular, the matter density does not decouple and the absence of a linear term in the correlation is notable. 

While this disagreement may not be an immediate problem given the current status of the data, one can expect that as data quality improves, these differences will become more pronounced. This may necessitate changing the assumed correlations (\ref{correlation}) or attempting to identify a corresponding field theory - assuming one exists - without a linear term in the expansion (\ref{xi_expand}). 

\section{Discussion}
Over two decades have passed since the discovery of dark energy \cite{Riess:1998cb, Perlmutter:1998np} and the possibility that dark energy evolves, as evident by the number of traditional models \cite{Cooray:1999da, Astier:2000as, Efstathiou:1999tm, Chevallier:2000qy, Linder:2002et, Jassal:2005qc,Barboza:2008rh}, has become a staple of cosmology.  Some of these parametrisations are attractive in the sense that they are the starting point of a Taylor expansion in a small parameter, either $z$ \cite{Cooray:1999da, Astier:2000as} or $(1-a)$ \cite{Chevallier:2000qy, Linder:2002et}. In contrast, others are more ad hoc \cite{Efstathiou:1999tm, Jassal:2005qc,Barboza:2008rh}.  Nevertheless, a common feature of all parametric models is that they are insensitive to DDE at $z \approx 0$ and only build up sensitivity with redshift. 

A key take-home message from this work is that the ``sensitivity" depends on the function multiplying the parameter $w_a$ in the usual two-parameter approach ($w_0, w_a$). Remarkably, the ubiquitous CPL model performs poorly and leads to larger errors on $w_a$, as should be evident from our mock analysis. In other words, if one is actively looking for evidence for DDE, ideally at the $\sim 3 \sigma$ level, then other models will reach that threshold first. Thus, assuming DDE is real, one may be confronted with a scenario where one has $\sim 1 \sigma$ in one parametric DDE model, but $\sim 3 \sigma$ in another. It is this arbitrariness, which is inherent in the parametric approach, that ultimately brings the parametric approach into question. In short, traditional DDE models are imperfect probes of DDE that are biased by the assumptions of the functional form of the $w_a$ term. Thus, it is prudent to test for DDE across a wide range of parametric models \cite{Yang:2021flj, Zheng:2021oeq}, especially at low redshift where the differences are especially acute. 

In the latter part of this work, we took a closer look at claims of oscillatory behaviour in $w(z)$ or $X(z)$ \cite{Zhao:2017cud, Wang:2018fng}. The rational for doing so is simply if the wiggles are real, then, as confirmed in the text, traditional parametric models will struggle to recover oscillatory behaviour. In the papers \cite{Zhao:2017cud, Wang:2018fng} what has been objectively shown is that working within the imposed correlations (\ref{correlation}), there is a range of parameters ($\sigma_m, a_c, \Delta_{X}$) where wiggles appear in reconstructions of $w(z)$ or $X(z)$. Further constraining this range, there are even values of $(\sigma_m, a_c, \Delta_{X})$ favoured by Bayesian evidence over flat $\Lambda$CDM. That being said, it should be obvious that all wiggles are observed through the prism of assumptions on correlations and different correlations interpret the \textit{same} data differently. 

So our motivation here was to strip the wiggles from the correlations employed in the reconstruction. To do so, we worked with the default $X(z)$ output from \cite{Wang:2018fng} and performed a DFT of $X(z)$. This analysis revealed only a finite number of relevant modes. This outcome may have been anticipated from the correlations (\ref{correlation}), since Fourier transform of the correlation corresponds to an exponential decay in frequency. Using the DFT output as a guide, we constructed wiggly dark energy models, which we confirmed through direct fits to the reconstructed $X(z)$. With these models, or ansatze, in hand, we further confronted them to the original data and recovered the wiggles from the data. It should be stressed that we dropped a local $H_0$ prior and replaced JLA supernovae with Pantheon supernovae. This suggests that the wiggles in the data may be robust enough. Of course, this vindicates the approach of \cite{Zhao:2017cud, Wang:2018fng} and raises further questions for parametric DDE models, since they cannot recover this behaviour.  

Finally, since the correlations employed in \cite{Zhao:2017cud, Wang:2018fng} do not appear to be physically motivated, but more motivated by the data, we attempted to interpret the correlations within a truncated Quintessence theory. The truncation is important here as it allows one to define the theory at low redshift, while relaxing the strict requirement that $ w(z) \geq -1$.  Within this framework, we noted in constrast to the assumptions made in \cite{Zhao:2017cud, Wang:2018fng},  it may not be possible to decouple the dark energy sector from other cosmological parameters, e.g. $\Omega_{m0}$. Moreover, when expanded around $z=0$, one sees that the correlation (\ref{correlation}) has no linear term. On the contrary, a linear term is expected from a generic field theory, as we show in fits to representative data. It is not clear if one can recover the assumed correlation (\ref{correlation}) from a field theory, but it is a very interesting question to explore, especially as the data quality improves.  

\section*{Acknowledgements}
We thank and Stephen Appleby, Shahab Joudaki, Eric Linder, Levon Pogosian \& Tao Yang for discussion/correspondence and comments on earlier drafts. We thank Yuting Wang \& Gong-Bo Zhao for sharing and explaining the $X(z)$ wiggles from \cite{Wang:2018fng}. E\'OC is funded by the National Research Foundation of Korea (NRF-2020R1A2C1102899). MMShJ would like to acknowledge SarAmadan grant No. ISEF/M/99131.
Lu Yin was supported by Basic Science Research Program through the National Research Foundation of Korea(NRF) funded by the Ministry of Education through the Center for Quantum Spacetime (CQUeST) of Sogang University (NRF-2020R1A6A1A03047877).

\appendix

\end{document}